\documentclass[prb,aps,twocolumn,showpacs,floatfix,superscriptaddress]{revtex4}
\usepackage{graphicx}
\usepackage{amsmath}
\usepackage{amssymb}

\begin{document}

\title{Transport properties of partially equilibrated quantum wires}

\author{Tobias Micklitz} 

\affiliation{Materials Science Division, Argonne National Laboratory,
 Argonne, Illinois 60439, USA}

\author{J\'er\^ome Rech} 

\affiliation{Materials Science Division, Argonne National Laboratory,
 Argonne, Illinois 60439, USA}

\affiliation{Physics Department, Arnold Sommerfeld Center for Theoretical
 Physics, and Center for NanoScience, Ludwig-Maximilians-Universit\"at,
 Theresienstrasse 37, 80333 Munich, Germany}

\author{K. A. Matveev} 

\affiliation{Materials Science Division, Argonne National Laboratory,
 Argonne, Illinois 60439, USA}

\date{\today} 


\begin{abstract}
 We study the effect of thermal equilibration on the transport properties
 of a weakly interacting one-dimensional electron system. Although
 equilibration is severely suppressed due to phase-space restrictions and
 conservation laws, it can lead to intriguing signatures in partially
 equilibrated quantum wires. We consider an ideal homogeneous quantum
 wire.  We find a finite temperature correction to the quantized
 conductance, which for a short wire scales with its length, but
 saturates to a length-independent value once the wire becomes
 exponentially long.  We also discuss thermoelectric properties of long
 quantum wires.  We show that the uniform quantum wire is a perfect
 thermoelectric refrigerator, approaching Carnot efficiency with
 increasing wire length.
\end{abstract}
\maketitle

\section{Introduction}

The quantization of the dc conductance in ballistic quantum wires, first
observed about two decades ago,\cite{cq1,cq2} is one of the fundamental
discoveries of mesoscopic physics.  The staircase-like dependence of the
conductance on the electron density, with plateaus at integral numbers of
$2e^2/h$ is readily understood from the single-electron
picture.\cite{landauer} The latter associates each plateau with a fixed
number of occupied electronic subbands, each supplying one quantum of
conductance $2e^2/h$.  On the other hand, interactions between
one-dimensional electrons often lead to qualitatively new phenomena.
These are commonly described within the so-called Luttinger liquid
theory,\cite{giamarchi} drastically different from Landau's Fermi liquid
description applicable to higher-dimensional systems.  The remarkable
success of the simple single-electron picture in describing the
quantization of conductance is attributed to the fact that quantum wires
are always connected to two-dimensional leads, where interactions between
electrons do not play a significant role.  In fact, it was shown in
Refs.~[\onlinecite{maslov,safi,ponomarenko}] that in an ideal Luttinger
liquid connected to Fermi liquid leads, the dc conductance is completely
controlled by the latter and, therefore, is not affected by interactions
in the wire.

For that reason, the discovery of small temperature-dependent deviations
from perfect quantization\cite{0.7-1,0.7-2,0.7-3,0.7-4, 0.7-5, 0.7-6,
 0.7-7,0.7-8,0.7-9} of the conductance of quantum wires at low electron
densities raised a lot of interest.  These generally manifest themselves
as a shoulder-like structure just below the first plateau of conductance.
Weak at the lowest temperatures available, this feature becomes more
significant as the temperature is increased, turning into a quasi-plateau
at about $0.7\times (2e^2/h)$.  A number of theoretical efforts trying to
reveal the microscopic mechanism of this so-called ``0.7 structure'' have
been made. Several spin-related approaches attribute the effect to
spontaneous polarization of the electron spins in the
wire\cite{0.7-1,spinpol1,spinpol2} or the existence of a local
spin-degenerate quasi-bound state playing the role of a Kondo
impurity.\cite{kondo1,kondo2} Other approaches discuss the role of
scattering from plasmons,\cite{plasmons} spin waves,\cite{spinwaves} or
phonons.\cite{phonons}

Despite the absence of a commonly accepted microscopic theory, it is
generally recognized that electron-electron interactions must be included
to account for the effect.  As a consequence, a number of recent
publications reconsider the effect of interactions on the transport
properties of one-dimensional conductors, going beyond the picture of an
ideal Luttinger liquid.\cite{jerome1, jerome2, jerome3, lunde1, lunde2,
 wang, spivak, bruus, tokura, meir, meidan, syljuasen, matveev1, mirlin}
Here we focus on a very fundamental aspect of interactions, studying how
they lead to the equilibration inside the wire of electrons coming from
the two leads.  We emphasize that this effect is absent in an ideal
Luttinger liquid.  Indeed, the bosonic elementary excitations of the
Luttinger liquid have infinite lifetime, thus there is no relaxation
towards equilibrium in these systems, no matter how strong the
interactions.  Within the Luttinger-liquid theory the processes leading to
the equilibration of the electron system would be accounted for by the
additional terms in the Hamiltonian, which are irrelevant in the
renormalization group sense.  Instead of pursuing this strategy, we
consider the regime of weakly interacting electrons, thereby avoiding the
complexity of the Luttinger-liquid picture.

\begin{figure}[tbp]
\centering
\resizebox{.42\textwidth}{!}{\includegraphics{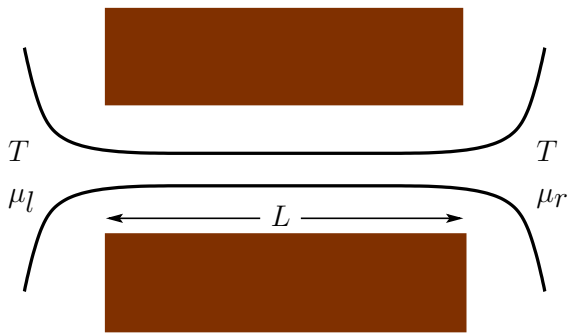}}
\caption{Schematic picture of the quantum wire of length $L$ which is
  formed by confining a two-dimensional electron gas with gates (dark
  regions). Electrons in the left and right lead are described by Fermi
  distribution functions characterized by temperature $T$ and chemical
  potentials $\mu_l$ and $\mu_r$, respectively.}
\label{fig1}
\end{figure}

Non-interacting electrons propagate ballistically through the wire and,
therefore, keep memory of the lead they originated from.  Thus the
distribution function of electrons inside the wire depends on the
direction of motion.  For the right- and left-moving particles it is
controlled, respectively, by the left and right lead:
\begin{equation}
\label{distributions}
f_p^{(0)} = \frac{\theta(p)}{e^{(\epsilon_p - \mu_{l})/T}+1}
          +\frac{\theta(-p)}{e^{(\epsilon_p - \mu_{r})/T}+1}.
\end{equation}
Here $\epsilon_p$ is the energy of an electron with momentum $p$ and
$\theta(p)$ is the unit step function.  The left and right leads are
assumed to have the same temperature $T$, but different chemical
potentials $\mu_l=\mu + eV$ and $\mu_r=\mu$ (see Fig.~\ref{fig1}).  Using
the distribution function (\ref{distributions}) one easily finds the
electric current $I=G_0V$, with the conductance
\begin{equation}
 \label{eq:G_0}
 G_0 = \frac{2e^2}{h}\frac{1}{e^{-\mu/T}+1},
\end{equation}
which coincides with the well-known conductance quantum $2e^2/h$ up to an
exponentially small correction \mbox{$\sim e^{-\mu/T}$}.

In the presence of interactions, the ballistic propagation of electrons
through the wire may be interrupted by collisions with other electrons.
As a result of these collisions, some electrons change their direction of
motion thus losing the memory of the lead they originated from.  Such
scattering processes modify the electron distribution function which is
then no longer given by Eq.~(\ref{distributions}).  The effect of the
electron-electron collisions on the distribution function depends strongly
on the length of the wire.  Indeed, electrons traverse short wires
relatively fast, so the interactions do not have the time to change
distribution (\ref{distributions}) considerably.  On the other hand, in
the limit of a very long wire one should expect full equilibration of
left- and right-moving electrons into a single distribution, even in the
case of weak interactions.

To simplify the subsequent discussion, in this paper we consider the case
of electrons with quadratic spectrum, $\epsilon_p=p^2/2m$, where $m$ is
the electron effective mass.  Then the system is Galilean invariant, and
one can easily infer the electron distribution function in the fully
equilibrated state.  Viewed from a frame moving with the drift velocity
$v_d=I/ne$ (where $I$ is the electric current and $n$ is the electron
density) the electron system is at rest and must be described by the
equilibrium Fermi distribution.  Performing a Galilean transformation back
into the stationary frame of reference this distribution takes the form
\begin{equation} 
\label{vddistrib}
f_p = \frac{1}{e^{ (\epsilon_p - v_d p - \bar\mu)/{\cal T}}+1}, 
\end{equation} 
where the chemical potential $\bar\mu$ and temperature ${\cal T}$
inside the equilibrated wire are, in general, different from $\mu_{l/r}$
and $T$.  At zero temperature, $T={\cal T}=0$, the distributions
(\ref{distributions}) and (\ref{vddistrib}) coincide, provided
$\mu_{l/r}=\bar\mu\pm v_d p_F$, where $p_F=\pi\hbar n/2$ is the Fermi
momentum of the system.  At non-zero temperature the distribution function
(\ref{vddistrib}) of electrons inside the wire is slightly different from
the distribution (\ref{distributions}) supplied by the leads.  In a
previous work\cite{jerome3} we have shown that the mismatch of the
distribution functions inside a very long wire and in the leads results in
additional contact resistance, reducing the conductance to
\begin{equation}
 \label{eq:conductance_RMM}
 G_\infty=\frac{2e^2}{h}\left[1-\frac{\pi^2}{12}
                         \left(\frac{T}{\mu}\right)^2\right].
\end{equation}
It is worth noting that the quadratic in $T/\mu$ correction in
Eq.~(\ref{eq:conductance_RMM}) is much more significant than the
exponentially small correction in Eq.~(\ref{eq:G_0}).

\begin{figure}[tb]
\centering
\resizebox{.48\textwidth}{!}{\includegraphics{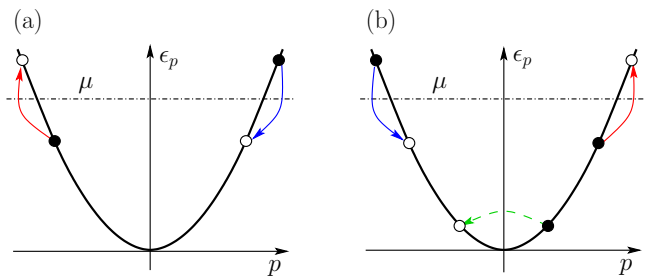}}
\caption{(a) Energy-conserving two-particle scattering process violates
 conservation of momentum.  Such processes can occur only in
 inhomogeneous systems.  (b) Dominant three-particle collision which
 gives rise to corrections to the conductance of short quantum
 wires.\cite{lunde1} A hole at the bottom of the band scatters off
 electron excitations close to the Fermi level.}
\label{fig2}
\end{figure}

The mechanism of equilibration of the electron distribution function in
one dimension is not fully understood.  While in higher dimensional
systems equilibration at low temperature is primarily provided by pair
collisions of electrons, these do not provide a relaxation mechanism in
one dimension.  This is due to the conservation laws for momentum and
energy which severely restrict the phase-space available for scattering
processes, Fig.~\ref{fig2}(a).  As a result, pair collisions in
one-dimensional wires can only occur with a zero momentum exchange or an
interchange of the two momenta, leaving the distribution function
unaffected.  The leading equilibration mechanism thus involves collisions
of more than two particles.  For a weakly interacting system, it is then
natural to assume that equilibration is provided by three-electron
scattering processes.

The effect of three-particle collisions on the transport properties of
short wires has been studied in a recent work by Lunde, Flensberg, and
Glazman.\cite{lunde1} In such short systems the effect of equilibration is
weak and the distribution function can be calculated perturbatively from
the distribution of non-interacting electrons (\ref{distributions}) within
the Boltzmann equation framework. Following this approach, Lunde {\it et
 al.}\cite{lunde1} obtained interaction-induced corrections to transport,
which they attributed to specific three-particle scattering events that
change the number of left- and right-movers. Indeed, in the absence of
interactions, the current flowing through the system can be viewed as the
superposition of the right- and left-moving flows of electrons supplied by
the left and right leads, respectively.  Once interactions are included,
these individual contributions change due to electron-electron collisions,
and one needs to account for the fact that electrons can now change
direction.  The electric current flowing through the wire is thus given by
the sum of the non-interacting part $I_0=G_0V$, and the change in, say,
the number of right-moving electrons inside the wire
\begin{equation}
\label{i1}
I=G_0V+ e\dot{N}^R.
\end{equation} 
Interaction-induced corrections to transport therefore arise from
processes which change the number of right- and left-moving electrons
rather than a change in the velocity of the charge carriers, as also
pointed out in Ref.~[\onlinecite{lunde1}].  

As shown by Lunde {\it et al.},\cite{lunde1} the most efficient
three-particle process changing the number of right-moving electrons
involves scattering of an electron into an empty state near the bottom of
the band, see Fig.~\ref{fig2}(b).  By calculating the resulting $\dot
N^R$, they obtained the correction to the conductance (\ref{eq:G_0}) of
the wire of the form
\begin{equation}
 \label{eq:deltaG_Lunde}
 \delta G = - \frac{2e^2}{h}\, \frac{L}{l_{eee}}\, e^{-\mu/T},
\end{equation}
where the length $l_{eee}$ is determined by the interaction strength and
shows a power-law temperature dependence.  The exponential smallness of
the correction (\ref{eq:deltaG_Lunde}) is due to the small probability of
finding an empty state near the bottom of the band.  Since the small
backscattering probability grows linearly with the length $L$, the
correction $\delta G\propto L$.

Because the same three-particle processes are responsible for the thermal
equilibration of the distribution (\ref{distributions}) into
(\ref{vddistrib}), the papers Ref.~[\onlinecite{jerome3}] and
Ref.~[\onlinecite{lunde1}] reviewed above study the same physical
phenomenon, albeit in the opposite limits of a long and a short wire.  In
the present paper we bridge the gap between these two limits.  We discuss
how the electron distribution evolves from the out-of-equilibrium form
(\ref{distributions}) in a short wire to a fully equilibrated form
(\ref{vddistrib}) in a long wire, and study how transport is affected by
the process of equilibration.  Our analysis focuses on weak
electron-electron interactions. It is thus formulated entirely in terms of
electrons, and does not use the bosonization technique.

The paper is organized as follows.  In sections~\ref{sec1} and~\ref{sec2}
we investigate how the conductance changes with increasing length of the
wire.  In Sec.~\ref{sec1} we expand on the kinetic-equation
treatment\cite{lunde1} of backscattering in short wires and study the
length dependence of the conductance while the correction $\delta G$
remains exponentially small.  In section~\ref{sec2} we turn to the regime
of exponentially long wires, where the correction $\delta G\sim (e^2/h)
(T/\mu)^2$, cf.~Eq.~(\ref{eq:conductance_RMM}).  In section~\ref{sec4} we
study the thermoelectric effects and show that the uniform quantum wire is
a perfect thermoelectric refrigerator, attaining Carnot efficiency with
increasing wire length.  Details of some calculations can be found in the
Appendices.

\section{Conductance of short wires}
\label{sec1}

Consider a quantum wire of length $L$, connected by ideal reflectionless
contacts to non-interacting leads biased by voltage $V$.  We are
interested in the process of thermal equilibration of the electrons inside
the wire, i.e., in how the transition from distribution (\ref{distributions})
to (\ref{vddistrib}) occurs, and how it affects the transport properties
of the system.

Following Lunde {\it et al.},\cite{lunde1} we describe the electron
transport in the wire in the framework of the Boltzmann equation
\begin{equation}
 \label{eq:Boltzmann}
 \frac{p}{m}\frac{\partial f_{p,x}}{\partial x} = I_{p,x}[f].
\end{equation}
We consider the steady-state setup in which the electron distribution
function $f_{p,x}$ depends on the position $x$ along the wire, but not on
time.  The collision integral $I_{p,x}[f]$ is, in general, a nonlinear
functional of the distribution function, whose form is determined by the
scattering processes affecting the distribution function.  As discussed
above, in our case the dominant processes are three-particle collisions,
in which case
\begin{widetext}
\begin{equation}
\label{collisionintegral}
I_{p_1,x}[f] 
= - \sum_{\footnotesize
 \begin{array}{l}
   p_2,p_3,\\ \sigma_2, \sigma_3
 \end{array}}
 \sum_{\footnotesize
 \begin{array}{l}
   p'_1,p'_2,p'_3\\ \sigma'_1,\sigma'_2,\sigma'_3
 \end{array}} w_{123;1'2'3'} \left[ f_1f_2f_3
\left(1-f_{1'}\right)\left(1-f_{2'}\right)\left(1-f_{3'}\right) 
       - f_{1'}f_{2'}f_{3'}\left(1-f_1\right)
\left(1-f_2\right)\left(1-f_3\right)\right],
\end{equation}
\end{widetext} 
where $w_{123;1'2'3'}$ is the rate for scattering the set of incoming
states $\{p_1\sigma_1,p_2\sigma_2,p_3\sigma_3\}$ into the set of outgoing
states $\{p'_1\sigma'_1,p'_2\sigma'_2,p'_3\sigma'_3\}$, and for notational
convenience we shortened $f_i=f_{p_i,x}$.

The Boltzmann equation (\ref{eq:Boltzmann}) should be solved with the
boundary conditions stating that the distributions $f_{p,0}$ of the
right-moving electrons ($p>0$) at the left end of the wire and $f_{p,L}$
of the left-moving electrons ($p<0$) at the right end coincide with the
distribution function $f_p^{(0)}$ in the leads, Eq.~(\ref{distributions}).
The conductance of the wire can then be found from Eq.~(\ref{i1}), with
the rate of change in the number of right-moving electrons related to the
collision integral via
\begin{equation}
\label{nr}
\dot{N}^R = 2 \int_0^ L dx \int_0^\infty {dp\over h} I_{p,x}[f].
\end{equation} 

Solving the Boltzmann equation exactly is a very difficult problem due to
the non-linearity of the collision integral (\ref{collisionintegral}), so
one generally has to make some simplifying assumptions.  Such assumption
in our case is that the temperature $T$ is small compared to the chemical
potential $\mu$.  

Clearly, at $T=0$ no real scattering processes are allowed, and the
unperturbed distribution $f_p=\theta(p_F-|p-mu|)$ solves the Boltzmann
equation (\ref{eq:Boltzmann}) for any value of the drift velocity $u$.
Since in this case the collision integral $I_{p,x}[f]$ vanishes, we get
$\dot N^R=0$, and, according to (\ref{i1}), the conductance of the wire is
$2e^2/h$.

A finite temperature $T$ acts in two important ways. First, it affects
states near the Fermi level: the step in the zero-$T$ distribution
softens, providing partially occupied states in a momentum range $\delta
p\sim T/v_F$ around the Fermi points.  Secondly, it ensures a finite
occupation of a hole (i.e., a vacant state) near the bottom of the band.
Although the occupation probability of such a hole is exponentially small,
$1-f_p\sim e^{-\mu/T}$, its presence is crucial for the three-particle
processes that change the number of right-moving electrons, see
Fig.~\ref{fig2}(b).  It is important to realize that the backscattering of
holes is accompanied by scattering of electrons near the Fermi points,
Fig.~\ref{fig2}(b).  In fact, this is the mechanism of the equilibration of
the distribution function to the form (\ref{vddistrib}) in long wires.
Although the backscattering rate is exponentially small, $\dot N^R\propto
e^{-\mu/T}$, it scales with the length of the wire.  Thus the full
equilibration is achieved in wires whose length $L$ exceeds an
exponentially long equilibration length $l_{\rm eq}\propto e^{\mu/T}$.
The exact definition of $l_{\rm eq}$ will be given below, see
Eq.~(\ref{leq}).

In this section we will discuss the case of short wires, $L\ll l_{eq}$.
The regime $L\gtrsim l_{\rm eq}$ will be discussed in Sec.~\ref{sec2}.

\subsection{Very short wires}
\label{pc}

We start our discussion with the case of very short wires, recently
considered by Lunde, Flensberg, and Glazman.\cite{lunde1} The authors
argued that for short enough wires, the interactions have little time to
change the distribution function from its initial value $f_p^{(0)}$ given
by Eq.~(\ref{distributions}), allowing one to perform a perturbative
expansion in the scattering rate $w_{123;1'2'3'}$.  In the lowest order,
this amounts to approximating the collision integral as
\begin{equation}
\label{approx0}
I_{p,x}[f] \simeq I_{p,x}[f^{(0)}].
\end{equation} 
Solving the Boltzmann equation to this approximation, they obtained an
expression for the modified distribution function inside the wire, which
they used to compute the electric current to first order in the scattering
rate.

The resulting correction to the conductance of the wire has the form
(\ref{eq:deltaG_Lunde}), in which microscopic details of the interaction
potential are absorbed into the length $l_{eee}$.  Lunde {\it et
 al.}\cite{lunde1} performed their calculation for a specific model of
electrons interacting via a potential defined by its Fourier transform
$V_q=V_0(1-q^2/q_0^2)$.  This expression results from the expansion of a
general potential under the assumption that small-momentum scattering is
dominant. The parameter $q_0\ll k_F$ accounts for the screening by the
nearby metallic gates, while $V_0$ is the zero-momentum Fourier component
of the screened Coulomb potential.  Within this model, the length
$l_{eee}$ is given by\cite{lunde1}
\begin{equation}
\label{lunde}
l_{eee}^{-1} \sim {\left(V_0 k_F\over \mu\right)^4} 
             \left({k_F\over q_0}\right)^4
             \left({T\over \mu}\right)^7  k_F.
\end{equation} 
A more careful treatment of the Coulomb interaction screened by a gate
leads to an additional logarithmic temperature dependence in
Eq.~(\ref{lunde}), see Appendix~\ref{aem}.

To better understand the result (\ref{eq:deltaG_Lunde}) and find the
limits of its applicability, we discuss the qualitative picture of this
phenomenon.  Let us focus on a single three-electron collision process.
The most favorable collision involves a maximal number of states close to
the Fermi points.  However, due to the conservation of both energy and
momentum, collisions that change the number of right- and left-movers
cannot occur near the Fermi level, and have to involve states deep in the
electron band. As pointed out by Lunde {\it et al.},\cite{lunde1} the
scattering process most susceptible to alter the current thus typically
scatters two electrons close to the Fermi points and one electron at the
bottom of the band, as schematically depicted in Fig.~\ref{fig2}(b). It is
convenient to think of this collision as a process in which a deep hole,
corresponding to the outgoing electron state, is backscattered by electron
excitations close to the Fermi level. These excitations are typically
associated with a momentum change $|\delta p|\sim T/v_F$ due to
Fermi-blocking, so that the backscattering occurs over a distance $\sim
T/v_F$ in momentum space. Let us furthermore characterize this process by
introducing a scattering rate $1/\tau_0$, which can be approximated by a
constant since the initial and final states both lie at the bottom of the
band.

The change $\dot{N}^R$ in the number of right-moving electrons per unit
time, due to these three-particle collisions can then be readily obtained.
It is given by the product of the scattering rate $1/\tau_0$ for one such
collision times the number of deep holes susceptible to be
backscattered. The latter can be estimated from the probability to find a
left- or right-moving hole $e^{-\mu^{L,R}/T}$ and the number of states
${(T/v_F)/(h/{ L})}$ available within the typical momentum range of
the backscattering process.  Taking into account that the scattering of a
left- or a right-moving hole both contribute to $\dot{N}^R$, but with a
different sign, one finally has
\begin{eqnarray}
\label{nr0qualitative}
\dot{N}^R&=&\frac{2}{\tau_0}
           \left(e^{-\mu^R/T} - e^{-\mu^L/T}\right) \frac{T L}{h v_F} 
\nonumber\\
& = & -{2 \over \tau_0} { \Delta\mu \over hv_F} e^{-\mu/T} L,
\end{eqnarray} 
where $\Delta\mu=\mu^R-\mu^L$, and we absorbed potential numerical
prefactors into the definition of $\tau_0$.  Throughout this paper we use
subscripts $l$ and $r$ to denote the left and right leads, whereas
superscripts $L$ and $R$ refer to the left- and right-moving electrons.
In short wires the chemical potentials of electrons are not significantly
affected by the scattering processes, so $\mu^R=\mu_l$, $\mu^L=\mu_r$, and
$\Delta \mu=eV$.

We then notice that according to Eq.~(\ref{i1}) the correction to the
conductance of the wire due to the backscattering processes is $\delta G =
e\dot N^R/V$.  As a result we recover the result (\ref{eq:deltaG_Lunde})
of Lunde {\it et al.},\cite{lunde1} provided
\begin{equation}
 \label{eq:tau_0}
 \tau_0\sim\frac{l_{eee}}{v_F}.
\end{equation}

The derivation of Eq.~(\ref{eq:deltaG_Lunde}) relied on the assumption
that the occupation probability of a deep hole is well described by the
distribution of non-interacting particles, or, alternatively, that one can
approximate the collision integral according to Eq.~(\ref{approx0}). This
approximation holds in cases where the hole typically scatters no more
than once during its propagation through the wire, and any transition
between subsystems of left- and right-movers occurs in a single collision.
One thus expects this result to be valid for wires shorter than the mean
free path of the hole $l_0$.  Since the typical momentum of a hole
contributing to $\dot N^R$ is of order $T/v_F$, we estimate $l_0\sim
T\tau_0/p_F$.  Substituting the estimate (\ref{eq:tau_0}), we obtain
\begin{equation}
\label{eq:mean_free_path}
l_0 \sim  \frac{T}{\mu}\, l_{eee}.
\end{equation}
For the particular model of the interaction potential used in
Ref.~[\onlinecite{lunde1}] we estimate
\begin{equation}
\label{invl0}
l_0^{-1} \sim {\left(V_0 k_F\over \mu\right)^4} 
             \left({k_F\over q_0}\right)^4
             \left({T\over \mu}\right)^6  k_F .
\end{equation} 
In wires longer than $l_0$, holes near the bottom of the band experience
multiple collisions while they propagate through the wire, the
distribution function deviate significantly from the unperturbed form
(\ref{distributions}), and the result (\ref{eq:deltaG_Lunde}) is no longer
applicable.

\subsection{Longer wires: $l_0\ll L\ll l_{\rm eq}$}
\label{int}

In wires longer than $l_0$ a typical hole near the bottom of the band is
scattered many times while traversing the wire.  Each collision changes
its momentum by a small amount $\delta p\sim T/v_F\ll p_F$, with a sign
that varies in a random fashion.  The hole thus performs a random walk in
momentum space.  This picture is analogous to the diffusion of a Brownian
particle in air.  In the latter case, the change of momentum of the
particle in each collision is small because its mass is much larger than
that of the air molecules.  Similarly to the case of Brownian motion, one
can use the small parameter $\delta p/p_F\sim T/\mu$ to bring the
collision integral of holes to a much simpler Fokker-Planck form
\begin{equation}
\label{approx1}
I_{p,x}[g] \simeq  - \frac{\partial}{\partial p} \left( A(p) g_{p,x} 
    -  {1\over 2} \frac{\partial}{\partial p}  [B(p) g_{p,x}] \right),
\end{equation} 
where we introduced the hole distribution $g_{p,x}=1-f_{p,x}$.  The
functions $A(p)$ and $B(p)$ entering Eq.~(\ref{approx1}) are model
specific.  In the case of three-electron collisions they can be determined
explicitly.  They depend on the three-particle scattering rate as well as
the electron distribution function in the vicinity of the Fermi level.
The latter can be assumed to be unperturbed by the collisions in the wire
as long as $L\ll l_{\rm eq}\propto e^{\mu/T}$.  The resulting derivation
of $A(p)$ and $B(p)$ can be found in Appendix~\ref{afp}; here we provide
order of magnitude estimates.

First we notice that $B(p)$ has the physical meaning of the diffusion
coefficient in momentum space, i.e., the typical momentum change of a hole
over time $t$ behaves as $(\Delta p)^2\sim Bt$.  Assuming as before that
the hole changes its momentum by $\pm T/v_F$ once during time $\tau_0$, we
conclude that $(\Delta p)^2\sim (T/v_F)^2 t/\tau_0$ for $t\gg \tau_0$.
Thus we estimate
\begin{equation}
 \label{eq:B_estimate}
 B\sim\frac{T^2}{v_F^2\tau_0}\sim \frac{T^2}{v_Fl_{eee}},
\end{equation}
where we used our earlier estimate (\ref{eq:tau_0}) of $\tau_0$, and the
microscopic expression for $l_{eee}$ is given by Eq.~(\ref{lunde}).

Although $B$ is a function of momentum $p$, the typical scale of the
variations of $B(p)$ is $p_F$.  Thus for the particle at the bottom of the
band one can approximate $B(p)$ by its value at $p=0$, which we will
denote as $B$.  Then $A(p)$ is easily obtained by noticing that the
collision integral (\ref{approx1}) has to vanish if the hole distribution
function takes an equilibrium Boltzmann form
\begin{equation}
 \label{eq:equlibrium_holes}
 g_{p,x}^{(0)}=e^{p^2/2mT}e^{-\mu/T}.
\end{equation}
This condition leads to the relation $A(p)=Bp/2mT$, which is also
confirmed explicitly in Appendix~\ref{afp}.  Using this result one easily
transforms the Boltzmann equation (\ref{eq:Boltzmann}) to the form
\begin{equation}
\frac{p}{m} \frac{\partial g_{p,x}}{\partial x}   
=\frac B2 \frac{\partial}{\partial p} \left( -\frac{p}{m T}\, g_{p,x} 
 + \frac{\partial g_{p,x}}{\partial p} \right).
\label{fullfp}
\end{equation}
The boundary conditions express the fact that the distributions of the
right-moving holes at the left end of the wire and that of left-moving
holes at the right end are controlled by the respective leads:
\begin{subequations}
 \label{eq:boundary_conditions_holes}
 \begin{eqnarray}
   \label{eq:bc_left_lead}
     g_{p,0} &=& e^{p^2/2mT}e^{-(\mu+eV)/T},\quad\mbox{for $p>0$},\\
   \label{eq:bc_right_lead}
     g_{p,L} &=& e^{p^2/2mT}e^{-\mu/T},\quad\mbox{for $p<0$}.
 \end{eqnarray}
\end{subequations}
Here we again assumed $\mu_r=\mu$ and $\mu_l=\mu+eV$.  The kinetic
equation in the form (\ref{fullfp}) is applicable only to exponentially
rare holes with $|p|\ll p_F$.  Thus the Fermi statistics of the holes is
irrelevant, and the boundary conditions on the distribution function have
the Boltzmann form.  Finally, combining our earlier results (\ref{i1}),
(\ref{nr}), and (\ref{approx1}), we express the correction to conductance
of the wire as
\begin{subequations}
 \label{eq:deltaG_holes}
 \begin{equation}
   \label{eq:deltaG_vs_NRdot}
   \delta G = \frac{e}{V}\,\dot N^R,
 \end{equation}
with
 \begin{equation}
 \label{eq:NRdot_holes}
   \dot N^R = \frac{B}{h}\int_0^L dx
            \left(\frac{\partial g_{p,x}}{\partial p}\right)_{p=0},
 \end{equation}
\end{subequations}
i.e., conductance is determined by the behavior of the distribution
function near $p=0$.

The solution of equation (\ref{fullfp}) with boundary conditions
(\ref{eq:boundary_conditions_holes}) shows two different regimes,
depending on the length of the wire.  In relatively short wires the effect
of hole scattering is weak, and to first approximation one can assume that
the distribution function $g_{p,x}$ does not depend on position $x$ and
coincides with the distribution (\ref{eq:boundary_conditions_holes})
provided by the leads.  This distribution is discontinuous at $p=0$,
namely $g_p\to e^{-\mu_{l,r}/T}$ at $p\to\pm0$.  To be more precise, one
should notice that the Fokker-Planck approximation applies to wires of
length in the range $l_0\ll L\ll l_{\rm eq}$.  At the lower end of this
range, $L\sim l_0$ the holes near the bottom of the band are scattered a few
times by the electrons near the Fermi level and change their momentum by
$\sim T/v_F$.  Thus in the center of the wire the discontinuity of the
distribution $g_p$ is smeared by $(\Delta p)_0\sim T/v_F$.  As the wire
length increases, the diffusion of holes in momentum space becomes more
pronounced, and at a certain length scale $l_1$ the smearing $\Delta p$
reaches a larger scale $(\Delta p)_1=(mT)^{1/2}$.  (Indeed, $(\Delta
p)_1/(\Delta p)_0\sim \sqrt{\mu/T}\gg1$.)  We shall consider the regimes
$L\ll l_1$ and $L\gg l_1$ separately, as different approximations can be
applied to the kinetic equation (\ref{fullfp}) in these two cases.  The
estimate for the length scale $l_1$ will be obtained below, see
Eq.~(\ref{eq:l_1}).

\subsubsection{Wires of length in the range $l_0\ll L\ll l_1$}

Let us present the hole distribution function as $g=g^{(0)} +\tilde g$,
where $g^{(0)}_{p,x}$ is the equilibrium distribution
(\ref{eq:equlibrium_holes}) and $\tilde g_{p,x}$ is the correction caused
by the applied bias $V$.  (At small bias we expect $\tilde g\propto V$.)
The distribution $g^{(0)}_{p,x}$, of course, satisfies the kinetic
equation (\ref{fullfp}).  Then, since equation (\ref{fullfp}) is linear in
$g$, it also fully applies to $\tilde g_{p,x}$.  It is important to note,
however, that the two terms in the right-hand side of this equation are
not of the same order of magnitude.  Indeed, at $p\sim \Delta p$ we have
$\partial \tilde g/\partial p\sim \tilde g/\Delta p\gg (p/mT)\tilde g$,
provided $\Delta p\ll (mT)^{1/2}$.  Thus the propagation of holes through
the wires of length $L$ in the range $l_0\ll L\ll l_1$ is described by the
simplified equation
\begin{equation}
 \label{eq:Kramers_tilde_g}
\frac{p}{m} \frac{\partial \tilde g_{p,x}}{\partial x}   
=\frac B2 \frac{\partial^2 \tilde g_{p,x}}{\partial p^2}.
\end{equation}

\begin{figure}[tb]
\resizebox{.48\textwidth}{!}{\includegraphics{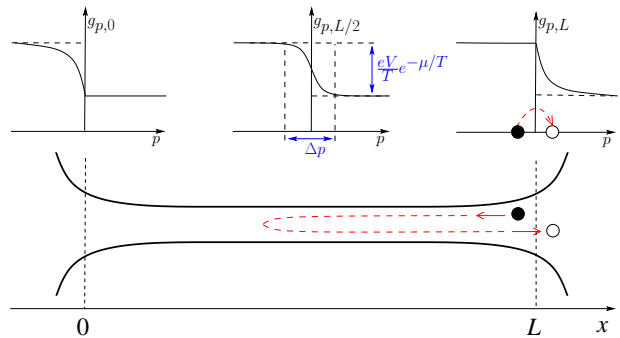}}  
\caption{As one goes along the wire from the left lead to the right one,
 the distribution function of holes changes.  The number of left-moving
 holes decreases as some of them turn around and start moving to the
 right.}
 \label{fig:holedistribution}
\end{figure}

To find the correction to conductance (\ref{eq:deltaG_holes}) for a wire
in the regime $l_0\ll L\ll l_1$ one needs to solve this equation with the
appropriate boundary conditions deduced from
Eq.~(\ref{eq:boundary_conditions_holes}).  We leave such a complete
solution for future work.  Instead, we perform a simple dimensional
analysis to conclude that the step in the distribution function near $p=0$
is broadened by 
\begin{equation}
 \label{eq:broadening}
 \Delta p\sim (BmL)^{1/3}.    
\end{equation}
This result can also be obtained from a simple physical argument.
Figure~\ref{fig:holedistribution} shows the hole distribution function
$g_{p,x}$ at different positions along the wire.  The scattering processes
contributing to the electric current involve holes entering from the right
lead with momentum $\Delta p$, moving to the left with their velocity
gradually decreasing, and eventually returning to the right lead.  In
order to lose the momentum of order $\Delta p$ the hole has to experience
sufficiently many collisions in the wire, which requires time $t$
determined from the standard diffusion condition $(\Delta p)^2\sim B t$.
Propagating through the wire at a typical velocity $\Delta p/m$ until the
turning point, the hole will move by distance $(\Delta p/m)t\sim L$.
Combining these two estimates, we recover our earlier result
(\ref{eq:broadening}).

At this point we can estimate the upper limit $l_1$ on the length of the
wire $L$, to which the approach used here is applicable.  In order to
neglect the term $B(\partial/\partial p)[(p/2mt)\tilde g]$ in the right
hand side of equation (\ref{eq:Kramers_tilde_g}) we assumed $\Delta
p\ll(mT)^{1/2}$.  From Eq.~(\ref{eq:broadening}) we see that this
approximation fails when the length of the wire reaches the value
\begin{equation}
 \label{eq:l_1}
 l_1\sim \frac{(mT^3)^{1/2}}{B}\sim \left(\frac{\mu}{T}\right)^{1/2} l_{eee},
\end{equation}
where we also used our earlier estimate (\ref{eq:B_estimate}) of $B$.  As
expected, $l_1\gg l_0$, see Eq.~(\ref{eq:mean_free_path}), i.e., the
approach used in this section applies to a parametrically broad range of
wire lengths.

To find the effect of three-particle scattering on the conductance of the
wire we use the boundary conditions (\ref{eq:boundary_conditions_holes})
to estimate
\[
\left(\frac{\partial g_{p,x}}{\partial p}\right)_{p=0}
\sim -\frac{1}{\Delta p} \frac{eV}{T} e^{-\mu/T}.
\]
Substituting this estimate into Eq.~(\ref{eq:deltaG_holes}), we obtain the
correction to the conductance in the form
\begin{equation}
\label{g2}
\delta G \sim -{2e^2\over h} \left({L \over l_1}\right)^{2/3} e^{-\mu/T}.
\end{equation} 
This correction should be compared with the result (\ref{eq:deltaG_Lunde})
of Lunde {\it et al.}\cite{lunde1} Both expressions are exponentially
small and grow with the length of the wire, but correction (\ref{g2})
shows a slower growth, $\delta G\propto L^{2/3}$, rather than linear
growth in Eq.~(\ref{eq:deltaG_Lunde}).  One can easily check that at the
crossover, $L\sim l_0$, the results (\ref{eq:deltaG_Lunde}) and (\ref{g2})
are of the same order of magnitude.

\subsubsection{Wires of length in the range $l_1\ll L\ll l_{\rm eq}$}

\label{sec:moderately_short}

\begin{figure}[tb]
\centering
\resizebox{.46\textwidth}{!}{\includegraphics{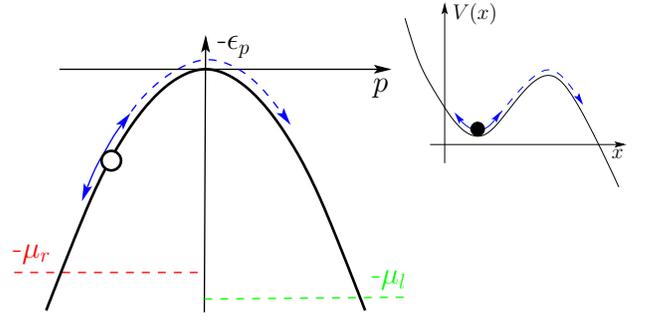}}
\caption{A hole with spectrum $-p^2/2m$  performs a
  random walk in momentum space in the presence of a barrier.
  Backscattering of a hole is analogous to a Brownian particle escaping
  from a local minimum of the potential, see inset.}
\label{passage}
\end{figure}

As mentioned above, a hole near the bottom of the band performs a random
walk in momentum space.  In the case of wires longer than $l_1$ one needs
to carefully consider the effect of the parabolic spectrum of the hole
$-p^2/2m$.  This spectrum plays the role of a potential barrier for the
random walker, see Fig.~\ref{passage}.  In order for the hole to
backscatter, and thus change sign of its momentum, it has to overcome the
barrier.  The rate of such backscattering events is controlled by the
height of the barrier measured from the Fermi level, and is exponentially
small as $e^{-\mu/T}$.  Evaluating the prefactor of backscattering rate is
an interesting problem, similar to that of a Brownian particle escaping
from a local minimum of the external potential.  The general features of
this problem are well understood.\cite{fpp} In order to overcome the
barrier the particle has to not just reach the top, but move beyond it far
enough for the potential to drop below the maximum by more than the
temperature $T$.  Applied to a hole diffusing in momentum space, this
means that the backscattering is controlled by the region of width $\Delta
p\sim \sqrt{mT}$ around $p=0$.

In wires shorter than $l_1$ this process cannot fully develop because of
the small time needed for the hole to traverse the wire, and one has to
include into consideration the spatial dependence of the distribution
function $g_{p,x}$.  At $L\gg l_1$ the holes spend enough time inside the
wire to fully complete the backscattering process.  Therefore, away from
the ends of the wire the distribution function no longer depends on
position.  As a result, the left-hand side of the kinetic equation
(\ref{fullfp}) vanishes, and it can be rewritten in the form
\begin{equation}
 \frac{\partial}{\partial p} \left( -\frac{p}{m T}\, g_{p} 
   + \frac{\partial g_{p}}{\partial p} \right)=0.
\label{eq:stationary_FP}
\end{equation}
To complete the mathematical formulation of the problem one has to impose
the appropriate boundary conditions on the distribution function.  Since
equation (\ref{eq:stationary_FP}) ignores the spatial dependence of the
distribution function, we cannot reuse the boundary conditions
(\ref{eq:boundary_conditions_holes}).  Instead we assume that the chemical
potentials of the right- and left moving holes are established by the
leads and do not vary along the wire:
\begin{equation}
\label{fpbc}
g_p =\left\{ 
\begin{array}{ll}
e^{p^2/2mT}e^{-(\mu + eV) /T}, & \text{for } p\gg\sqrt{mT},
\\
e^{p^2/2mT}e^{-\mu/T} ,&  \text{for }- p\gg\sqrt{mT}.
\end{array}
\right.
\end{equation} 
This assumption implies that the total number of backscattered holes is
too small to affect the chemical potentials.  This is justified by the
fact that the backscattering rate is exponentially small.  In wires of
exponentially large length $L\gtrsim l_{\rm eq}$ this condition is
violated.  The latter regime will be discussed in Sec.~\ref{sec2}.

The solution of equation (\ref{eq:stationary_FP}) with boundary conditions
(\ref{fpbc}) is straightforward and gives
\begin{eqnarray}
\label{result1}
g_p &=& e^{p^2/2mT}e^{-\mu/T}
\nonumber\\
    &&\times\left(1 +  \frac{e^{-eV/T}-1}{\sqrt{2\pi mT}}
                  \int_{-\infty}^p  e^{-{p'^2}/2mT}dp'  \right).
\end{eqnarray} 
This distribution function smoothly interpolates between the boundary
conditions (\ref{fpbc}) imposed by the applied bias.  As expected, the
crossover occurs in a narrow region of width $\Delta p \sim \sqrt{mT}$
at the bottom of the band.

To linear order in $eV/T$ we find
\begin{equation}
\label{result2}
\left(\frac{\partial g_{p}}{\partial p}\right)_{p=0}
= -\frac{eV}{\sqrt{2\pi mT^3}}\,  e^{-\mu/T},
\end{equation} 
resulting in the backscattering rate 
\begin{equation}
 \label{eq:NRdot_holes_result}
 \dot N^R = -\frac{eVBL}{h\sqrt{2\pi mT^3}}\,  e^{-\mu/T},
\end{equation}
see Eq.~(\ref{eq:NRdot_holes}).  As a result, the correction to the
conductance (\ref{eq:deltaG_vs_NRdot}) takes the form
\begin{equation}
\label{g1}
\delta G =  -{2e^2\over h} {L \over l_1} \,e^{-\mu/T},
\end{equation}  
where we have used the following precise definition 
\begin{equation}
 \label{eq:l_1_precise}
 l_1=\frac{\sqrt{8\pi mT^3}}{B}
\end{equation}
of the length $l_1$, consistent with our earlier estimate (\ref{eq:l_1}).
It is worth mentioning that for wires of length $L\sim l_1$ the
expressions (\ref{g2}) and (\ref{g1}) give the same estimate for $\delta
G$.

Our result (\ref{g1}) has the form similar to the prediction
(\ref{eq:deltaG_Lunde}) of Lunde {\it et al.}\cite{lunde1} for short
wires, $L\ll l_0$.  Both expressions for the correction to the conductance
are exponentially small, but grow linearly with the length of the wire
$L$.  However, due to the sublinear growth (\ref{g2}) in the intermediate
range of wire lengths $l_0\ll L \ll l_1$, the prefactor $l_1^{-1}$ in
Eq.~(\ref{g1}) is parametrically smaller than $l_{eee}^{-1}$ in
Eq.~(\ref{eq:deltaG_Lunde}), see (\ref{eq:l_1}).

The result (\ref{g1}) can be derived qualitatively, following arguments
similar to the ones used in Sec~\ref{pc}.  There the change $\dot{N}^R$
in the number of right-moving electrons per unit time was estimated as the
ratio of the number of holes likely to backscatter and the average time
$\tau_0$ of such backscattering event, see Eq.~(\ref{nr0qualitative}).
Compared to the case of very short wires considered in Sec~\ref{pc}, for a
hole to change direction, it must now cover a larger distance $(\Delta
p)_1 \sim\sqrt{mT}\gg (\Delta p)_0\sim T/v_F$ in momentum space set by the
smearing of the discontinuity of the distribution function at the bottom
of the band.  The number of states available for the passage is thus
larger by a factor $(\Delta p)_1/(\Delta p)_0\sim\sqrt{\mu/T}$ compared to
the case of a very short wire, Sec~\ref{pc}.  On the other hand, even
though the typical time between two three-particle collisions is still
given by $\tau_0$, it now takes many such collisions for a hole to
complete the backscattering process.  Because the hole performs a random
walk in momentum space, the time $\tau_1$ it takes to cover the longer
distance $(\Delta p)_1$ can be estimated from $\tau_1/\tau_0\sim(\Delta
p)_1^2/ (\Delta p)_0^2\sim \mu/T$.  Combining both effects we find that
the correction to the conductance (\ref{g1}) should be smaller than
(\ref{eq:deltaG_Lunde}) by a factor of $\sqrt{\mu/T}$, in agreement with
Eq.~(\ref{eq:l_1}).

\section{Conductance of long wires}
\label{sec2}

In short quantum wires, the distribution function of electrons remains
close to the unperturbed form (\ref{distributions}) provided by the leads.
The main change due to the processes of electron collisions occurs near
the bottom of the band, with the discontinuity at $p=0$ being gradually
smeared as the wire length $L$ increases.  Because the discontinuity
affects only electrons deep below the Fermi level, the effect of
collisions is exponentially small.  In particular, this enabled us to
neglect the effect of collisions on the chemical potentials and assume
that to first approximation the right- and left-moving electrons remain in
equilibrium with the left and right leads, respectively.

A much more significant change occurs in long wires, $L\sim l_{\rm
 eq}\propto e^{\mu/T}$, for which the exponential suppression of the
equilibration effects is compensated by a large system size.  Once the
length of the wire becomes exponentially large, the relaxation of the
electron system becomes significant, and eventually, at $L\gg l_{\rm eq}$
the distribution function assumes the fully equilibrated form of
Eq.~(\ref{vddistrib}).  Unlike the relatively minor modification of the
distribution function in short wires, the difference between the
distributions (\ref{distributions}) and (\ref{vddistrib}) is not
exponentially small and, more importantly, concentrated near the Fermi
points, rather than at the bottom of the band.  In this section we
consider the conductance of the partially equilibrated wires, of length
$L\sim l_{\rm eq}$.  We start by discussing the form of the electron
distribution in this regime.

\subsection{Electron distribution function in the case of partial
 equilibration} 

\label{sec:partial_distribution}

Let us consider a segment of the wire, whose length $\Delta L$ is small
compared to the equilibration length $l_{\rm eq}\propto e^{\mu/T}$.  This
condition implies that a typical electron with energy near the Fermi level
passes through the segment without backscattering.  On the other hand,
$\Delta L$ is assumed to be sufficiently large for electrons to experience
multiple three-particle collisions, which do not result in backscattering.
Under these circumstances, the electron distribution function in the
segment will achieve a state of partial equilibration, in which the
numbers $N^R$ and $N^L$ of the right- and left-moving electrons are not
changed by collisions.  The form of this distribution can be obtained from
a general statistical mechanics argument.  The multiple collisions
occurring in the system will maximize the entropy of the non-interacting
electrons
\begin{equation}
 \label{eq:entropy}
 {\cal S}=-2\sum_p[f_p\ln f_p +(1-f_p)\ln(1-f_p)],
\end{equation}
while preserving the total energy, momentum, $N^R$, and $N^L$, given by
\begin{subequations}
 \label{eq:integrals_of_motion}
\begin{eqnarray}
 E&=&2\sum_p \epsilon_p f_p,
 \label{eq:integral_energy}
\\
 P&=&2\sum_p p f_p,
 \label{eq:integral_momentum}
\\
 N^R&=&2\sum_{p>0}f_p,
 \label{eq:integral_N^R}
\\
 N^L&=&2\sum_{p<0}f_p.
 \label{eq:integral_N^L}
\end{eqnarray}
\end{subequations}
Subtracting from the functional (\ref{eq:entropy}) the expressions for the
conserved quantities (\ref{eq:integral_energy})--(\ref{eq:integral_N^L})
with the Lagrange multipliers $\beta$, $-\beta u$, $-\beta\mu^R$, and
$-\beta\mu^L$, respectively, and differentiating with respect to $f_p$, we
find that the maximum of entropy is achieved for the distribution
\begin{equation}
\label{dfus}
f_p = \frac{\theta(p)}{e^{(\epsilon_p-up-\mu^R)/{\cal T}}+1}
     +\frac{\theta(-p)}{e^{(\epsilon_p-up-\mu^L)/{\cal T}}+1}.
\end{equation}
Here ${\cal T}=1/\beta$ is the effective temperature, parameter $u$ has
dimension of velocity and accounts for conservation of momentum in
electron collisions, $\mu^R$ and $\mu^L$ are the chemical potentials of
the right- and left-moving particles.

It is worth mentioning that the distribution (\ref{dfus}) does not apply
to particles near the bottom of the band, $p\sim \sqrt{mT}$.  Indeed, for
a hole near $p=0$ collisions with and without backscattering (i.e., the
change of the sign of $p$) are roughly equally likely.  Thus the above
discussion is not applicable in this case.  In order to find the form of
the distribution function near the bottom of the band, one should perform
an analysis similar to that of Sec.~\ref{sec:moderately_short}.  In
particular, the exponentially small discontinuity of the distribution
(\ref{dfus}) at $p=0$ will be smeared.  On the other hand, most quantities
of interest are determined by the behavior of the distribution function
near the Fermi level.  For instance, using (\ref{dfus}) we obtain the
electric current in the form
\begin{equation}
\label{currentudmu}
I = {2e\over h}\Delta \mu + e nu ,
\end{equation} 
up to corrections small as $e^{-\mu/T}$.  Here $\Delta \mu=\mu^R-\mu^L$.

It is instructive to see how the distribution (\ref{dfus}) interpolates
between the regimes of no equilibration (\ref{distributions}) and that of
full equilibration (\ref{vddistrib}).  The unperturbed distribution
(\ref{distributions}) is obtained from (\ref{dfus}) by setting $u=0$ and
identifying the chemical potentials with those in the leads: $\mu^R=\mu_l$
and $\mu^L=\mu_r$.  In this case $\Delta\mu=eV$, and
Eq.~(\ref{currentudmu}) reproduces the Landauer formula.  The fully
equilibrated distribution (\ref{vddistrib}) is obtained from (\ref{dfus})
by setting $\Delta\mu=\mu^R-\mu^L=0$.  In this case the electric current
(\ref{currentudmu}) is expressed as $I=enu$, which identifies parameter
$u$ with the drift velocity $v_d$.

In the regime when the distribution function (\ref{dfus}) differs from the
limiting cases (\ref{distributions}) and (\ref{vddistrib}) it is
convenient to quantify the degree of equilibration in the wire by the
parameter 
\begin{equation}
\label{defeta}
\eta ={u\over v_d}.
\end{equation} 
The case of no equilibration corresponds to $\eta=0$ and that of full
equilibration to $\eta=1$.

The meaning of the distribution function (\ref{dfus}) can be further
clarified by considering the Boltzmann equation (\ref{eq:Boltzmann}).  The
scattering processes contributing into the collision integral
(\ref{collisionintegral}) fall into two categories.  The strongest
processes preserve the numbers of the right- and left-moving electrons,
whereas the ones resulting in backscattering are exponentially weak, as
discussed by Lunde {\it et al.}\cite{lunde1} and also above in
Sec.~\ref{sec1}.  Let us approximate the collision integral
(\ref{collisionintegral}) by neglecting the weak backscattering processes.
Then, by substituting the distribution (\ref{dfus}) into the right-hand
side of Eq.~(\ref{collisionintegral}), one easily sees that each term in
the sum vanishes.  Thus the distribution (\ref{dfus}) solves the Boltzmann
equation (\ref{eq:Boltzmann}) in this approximation.  Furthermore, in the
absence of backscattering the solution (\ref{dfus}) applies for any choice
of parameters ${\cal T}$, $u$, $\mu^R$, and $\mu^L$, and in particular,
for any degree of equilibration $\eta$.  The value of $\eta$ is ultimately
determined by the exponentially weak backscattering processes and the
length of the wire.

\subsection{Conservation laws} 
\label{cv}

Conductance of a long quantum wire, in which the electron distribution
function is fully equilibrated, was studied in
Ref.~[\onlinecite{jerome3}], where a power-law correction to the quantized
conductance was obtained, Eq.~(\ref{eq:conductance_RMM}).  The derivation
of this result was based on an analysis of conservation laws for the
number of electrons, energy, and momentum satisfied in electron
collisions.  Here we perform a similar analysis for a partially
equilibrated wire.

Conservation of the total number of particles $N$ implies that in a steady
state the particle current $j(x)$ is uniform along the wire.
Correspondingly, we infer from the conservation of total momentum $P$ and
total energy $E$ that in the steady state a constant momentum current
$j_P$ and a constant energy current $j_E$ flow through the system.  In the
following it will be convenient to express these currents as the sum of
the individual contributions from left- and right-moving electrons,
e.g. $j=j^R+j^L$, thus introducing
\begin{subequations}
\label{currents}
 \begin{eqnarray}
 \label{particle_currents}
j^{R/L}(x) &=& \int_{-\infty}^\infty {dp\over h} 
              \theta(\pm p) v_p  f_{p,x},
\\
 \label{momentum_currents}
j_P^{R/L}(x) &=& \int_{-\infty}^\infty {dp\over h} 
               \theta(\pm p) v_p p  f_{p,},
\\
 \label{energy_currents}
j_E^{R/L}(x) &=& \int_{-\infty}^\infty {dp\over h} 
               \theta(\pm p) v_p  \epsilon_p f_{p,x}. 
 \end{eqnarray}
\end{subequations} 
Here $v_p=p/m$ is the electron velocity, the positive sign in the step
function corresponds to right-movers, while the negative one to
left-movers.  

Near the ends of the wire, the distribution function of incoming electrons
is controlled by the leads. Close to the left lead, the distribution $f^R$
of right-moving electrons thus assumes the form of the first term in
Eq.~(\ref{distributions}), and similarly, close to the right lead, the
left-movers' distribution $f^L$ is given by the second term in
Eq.~(\ref{distributions}).  This allows us to readily calculate, for
example, the current $j^R(l)$ of right-moving electrons near the left end
of the wire. Unlike the total current $j$, the current $j^R(x)$ is not
uniform throughout the system, since the equilibration processes ensure
the conversion of right-moving electrons into left-moving ones.  From
conservation of the number of particles it follows that the total number
of right-moving electrons changing direction per unit time equals the
difference between their outgoing and incoming flows at the right and left
leads
\begin{equation}
\label{conteq}
\dot{N}^R = j^R (r) - j^R (l).
\end{equation} 
A calculation of $j^R(r)$ requires the knowledge of the distribution of
right-moving electrons at the right end of the wire.  As the latter is
unknown, we proceed by expressing $j^R(r)$ in terms of the total particle
current $j$ and the incoming flow of left-movers supplied by the right
lead
\begin{equation}
\label{ccurr}
j^R(r)= j - j^L(r),
\end{equation} 
where $j^L(r)$ can now be determined from the known electron distribution
in the right lead.  Combining Eqs.~(\ref{conteq}) and (\ref{ccurr}) we can
relate $\dot{N}^R$ to the total incoming flow of particles as
\begin{equation}
\label{jl+jr}
j^R(l) + j^L(r) =j-\dot{N}^R.
\end{equation} 
Using the lead distribution function (\ref{distributions}) the left-hand
side of (\ref{jl+jr}) is readily calculated, and takes the form $G_0V/e$,
with the conductance $G_0$ defined by Eq.~(\ref{eq:G_0}).  Noticing that
the electric current $I=ej$, we then recover Eq.~(\ref{i1}).  For the
purposes of this section we do not need to keep the exponentially small
corrections to $G_0$, and upon substitution $G_0=2e^2/h$ we are left with
the relation 
\begin{equation}
\label{j2}
{2e\over h} V = j - \dot{N}^R ,
\end{equation} 
between voltage, current, and the rate of change of the number of right
moving electrons due to collisions.

Let us now analyze the consequences of energy conservation in
electron-electron collisions.  Repeating the above steps for the energy
current $j_E$, we arrive at an expression
\begin{equation}
\label{eq:energy_conservation}
j_E^R(l) + j_E^L(r) =j_E-\dot{E}^R,
\end{equation} 
analogous to Eq.~(\ref{jl+jr}).  Here $\dot{E}^R$ is the rate of change of
the energy of right-movers.

The conservation of the number of electrons and energy ensure that the
currents $j$ and $j_E$ are constant along the wire.  It is convenient to
combine them into the heat current
\begin{equation}
\label{jqjej}
j_Q = j_E - \mu j, 
\end{equation} 
which is consequently also independent of position.\cite{footnote}
Combining Eq.~(\ref{eq:energy_conservation}) and (\ref{jl+jr}) we find
\begin{equation}
\label{eq:heat_conservation}
j_Q^R(l) + j_Q^L(r) =j_Q-\dot{Q}^R,
\end{equation} 
where 
\begin{equation}
 \label{eq:Q^Rdot}
 \dot{Q}^R = \dot{E}^R-\mu \dot{N}^R  
\end{equation}
is the heat transferred into the right-moving subsystem by electron
collisions. 

The left-hand side of Eq.~(\ref{eq:heat_conservation}) is the heat current
in a non-interacting quantum wire.  Direct calculation shows that it is
exponentially small (see also the discussion in the beginning of
Sec.~\ref{sec4}), and for our purposes here can be assumed to vanish.
We therefore conclude 
\begin{equation}
\label{hc}
j_Q=\dot{Q}^R.
\end{equation}

Since the heat current $j_Q$ does not depend on position, it can be
calculated at any point in the wire.  In the regions not too close to the
leads the distribution function is expected to have the partially
equilibrated form (\ref{dfus}).  Then, using the expressions
(\ref{particle_currents}) and (\ref{energy_currents}) for $j$ and $j_E$,
we obtain
\begin{equation}
\label{jq}
j_Q={\pi^2\over 6}{T^2\over\mu}nu
\end{equation} 
to leading order in $T/\mu$.  As expected, in the absence of
equilibration, $u=0$, the heat current vanishes.

In Sec.~\ref{sec:partial_distribution} we introduced the distribution
function (\ref{dfus}) by discussing a short segment of the wire.  The four
parameters of this distribution ${\cal T}$, $u$, $\mu^R$, and $\mu^L$ may,
in principle, vary along the wire.  The independence of heat current
(\ref{jq}) on position then shows that the velocity $u$ and, therefore,
the degree of equilibration $\eta$ are constant along the wire.
Furthermore, since the electric current (\ref{currentudmu}) and $u$ are
constant along the wire, one concludes that the difference of the chemical
potentials $\Delta \mu=\mu^R-\mu^L$ is constant as well.  The only two
parameters of the distribution (\ref{dfus}) that can vary along the wire
are the temperature ${\cal T}$ and the average chemical potential
$(\mu^R+\mu^L)/2$.  Their dependences on position are discussed in
Appendix~\ref{amdf}.

\subsection{Relation between the degree of equilibration $\eta$ and the
 conductance of the wire}
\label{cond}

To make further progress we elaborate on the relationship between the
rates $\dot{N}^R$ and $\dot{E}^R$, whose explicit forms depend on the
details of the equilibration mechanism.  As we discussed in
Sec.~\ref{sec:partial_distribution}, in the absence of scattering
processes which change the number of left- and right-moving electrons, the
distribution (\ref{dfus}) is unaffected by electron-electron collisions,
i.e., the collision integral (\ref{collisionintegral}) vanishes.  In
particular, for $u=0$ the unperturbed distribution (\ref{distributions})
supplied by the leads would retain its form in the wire.  The
backscattering processes, which by definition contribute to $\dot N^R$,
also change the energy of the subsystem of right-moving electrons,
resulting in a non-vanishing $\dot E^R$.  Because both rates are caused by
the same backscattering processes, one expects to find a relation between
$\dot N^R$ and $\dot E^R$.  Here we establish such a relation with the
help of conservation laws.  An alternative and more formal derivation can
be found in Appendix \ref{aernr}.

\begin{figure}[tb]
\centering
\resizebox{.4\textwidth}{!}{\includegraphics{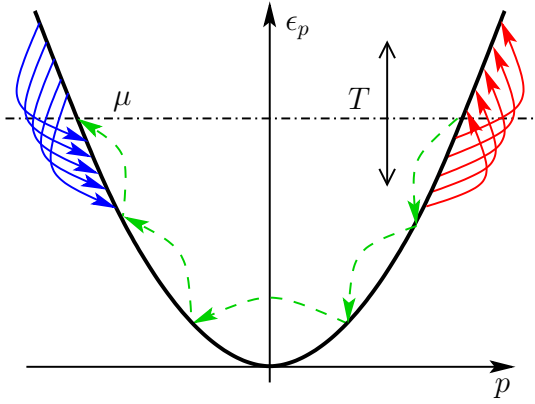}}
\caption{Sequence of elementary scattering events leading to the transfer
 of one electron from the right to the left Fermi point.  Conservation of
 momentum leads to the heating of the right-moving electrons and cooling
 of the left-moving ones.}
\label{multi3p}
\end{figure}

The backscattering processes transform the unperturbed distribution
(\ref{distributions}) to the partially equilibrated form (\ref{dfus}) with
non-vanishing $u$.  The two distributions differ most prominently at
energies within $\sim T$ of the Fermi level.  One can thus assume that all
the right-moving electrons contributing to $\dot N^R$ are removed from the
vicinity of the right Fermi point and placed to the vicinity of the left
one.  Each such transfer reduces the momentum of the system by $2p_F$.
Since the electron-electron collisions conserve momentum, a number of
other electrons have to be scattered in the vicinities of the two Fermi
points, see Fig.~\ref{multi3p}.  In the special case of three particle
collisions, the transfer of electron from the right Fermi point to the
left one is accomplished in a number of small steps with momentum change
$\delta p\sim T/v_F$, and at each step one additional electron is scattered
near each of the two Fermi points, see Fig.~\ref{fig2}(b).

As a result of the rearrangement of electrons near the two Fermi points,
the momentum change $2p_F$ of the backscattered electrons is distributed
between the remaining right- and left-moving electrons, i.e., $\Delta
p^R+\Delta p^L=2p_F$.  Thus the energy of the remaining right-movers
increases by $\Delta Q^R=v_F \Delta p^R$ whereas that of the left movers
decreases, $\Delta Q^L=-v_F \Delta p^L$.  Then, the conservation of energy
requires $\Delta p^R=\Delta p^L=p_F$.  In the end, the energy balance for
the right-moving electrons consists of a loss of $\epsilon_F$ due to
removal of one particle from the Fermi level and a gain of $\Delta Q^R=v_F
p_F=2\epsilon_F$ due to the redistribution of momentum.  As a result, for
every right-moving electron that changes direction, $\Delta N^R=-1$, the
right-movers' energy increases by an amount $\Delta E^R=\epsilon_F$.  It
is easy to check that the difference between the chemical potential $\mu$
and the Fermi energy $\epsilon_F$ is irrelevant for our discussion, so we
conclude
\begin{equation}
\label{ernr}
\dot{E}^R= -\mu\dot{N}^R.
\end{equation}  
It is important to point out that this result is independent of a specific
equilibration mechanism, or the degree to which equilibration has
occurred.  

The result (\ref{ernr}) can be also expressed in the form 
\begin{equation}
 \label{eq:QRdot-NRdot}
 \dot Q^R=-2\mu\dot N^R, 
\end{equation}
cf.~Eq.~(\ref{eq:Q^Rdot}), which expresses the simple fact that
when a right-moving electron is moved to the left Fermi point, the
remaining right-movers gain energy, see Fig.~\ref{multi3p}.  Combining
this expression with Eqs.~(\ref{hc}) and (\ref{jq}), we obtain
\begin{equation}
\label{nri}
\dot{N}^R = -r_0 nu,
\end{equation} 
where the dimensionless parameter $r_0$ is defined as
\begin{equation}
\label{r0}
r_0 ={\pi^2\over 12} \left({T\over \mu}\right)^2.
\end{equation} 
Noticing that by definition of the degree of equilibration (\ref{defeta})
$nu=\eta n v_d=\eta j$ and using the conservation of the particle number
in electron collisions, Eq.~(\ref{j2}), we find a linear relation between
the applied bias and the electric current flowing through the wire.  We
can thus readily extract the expression for the conductance in partially
equilibrated wires
\begin{equation}
\label{g}
G= {2e^2\over h}
  \left[ 1- \eta{\pi^2\over 12} \left({T\over \mu}\right)^2 \right],
\end{equation} 
where we discarded higher-order corrections in $(T/ \mu)^2$.  

The result (\ref{g}) reaffirms that at finite temperature the processes of
equilibration of the electron distribution function lead to a deviation of
the conductance from its quantized value $2e^2/h$.  In a fully
equilibrated wire $\eta \to 1$, and this correction saturates at a value
that does not depend on the details of the electron-electron interaction,
reproducing our earlier result (\ref{eq:conductance_RMM}).  The correction
to the conductance in this situation is quadratic in temperature $\delta G
\propto \left(T/\mu\right)^2$, in contrast with the results for the short
wire, where $\delta G \sim e^{-\mu/T}$.  

We obtained the expression (\ref{g}) for the conductance of the wire by
taking advantage of the conservation laws for the electron-electron
collisions as well as the basic properties of the non-interacting leads
the wire is connected to.  Although the expression (\ref{g}) is thus very
general, it does not fully determine the conductance of the wire, as the
parameter $\eta$ cannot be obtained in this approach, with the exception
of special cases of non-interacting electrons, $\eta=0$, and a very long
wire, $\eta=1$.  To find $\eta$ and, therefore, the conductance of the
wire, for arbitrary wire length, one needs to consider a specific model of
electron-electron interactions.  We now turn to such a calculation for the
most relevant case of relaxation via three-particle collisions.

\subsection{Partially equilibrated wires and equilibration length}

Our expression (\ref{g}) for the conductance of partially equilibrated
wires relied on the relation (\ref{nri}) between the rate $\dot N^R$ of
backscattering of right-moving electrons in the wire and the parameter $u$
of the distribution function (\ref{dfus}).  Another relation between $\dot
N^R$ and the distribution function can be found by considering the
microscopic mechanism of such backscattering, using the approach of
Sec.~\ref{sec:moderately_short}.  Comparison of the two expressions will
enable us to determine the degree of equilibration $\eta$ for wires of
arbitrary length. 

The form of the electron distribution in a partially equilibrated state
(\ref{dfus}) is controlled by four parameters, namely, the temperature
${\cal T}$, average chemical potential $\mu=(\mu^R+\mu^L)/2$, difference
of the chemical potentials $\Delta \mu=\mu^R-\mu^L$, and the velocity $u$.
In the absence of bias $V$ applied to the wire, the temperature ${\cal T}$
and chemical potential $\mu$ are equal to those in the leads, whereas
$\Delta\mu$ and $u$ vanish.  As a result the distribution (\ref{dfus})
reproduces the equilibrium Fermi-Dirac distribution, and clearly $\dot
N^R=0$.  Our goal is to find $\dot N^R$ to linear order in $V$.  

Applied bias affects the four parameters of the distribution (\ref{dfus})
differently.  As we discussed at the end of Sec.~\ref{cv}, the temperature
${\cal T}$ and chemical potential $\mu$ acquire position dependence,
whereas $\Delta\mu$ and $u$ no longer vanish, but remain constant along
the wire.  Thus, to linear order in $V$ one expects to find for the rate
$\dot n^R$ of backscattering per unit length of the wire
\begin{equation}
\frac{\dot N^R}{L} = \gamma_1 \Delta \mu +\gamma_2 u+\gamma_3\,\partial_x{\cal
 T}+\gamma_4\partial_x\mu.
\label{eq:linear_response}
\end{equation}
Because the gradients $\partial_x{\cal T}$ and $\partial_x\mu$ are caused
by bias applied to a long wire, they are not only proportional to $V$, but
also scale as $1/L$ (see also Appendix~\ref{amdf}).  Thus for the
exponentially long wires considered here, $L\sim l_{\rm eq}$, effect of
the gradients of ${\cal T}$ and $\mu$ can be neglected.  It is also clear
that $\gamma_2=0$.  Indeed, at $\Delta\mu=0$ the distribution takes the
fully equilibrated form (\ref{vddistrib}), for which no relaxation takes
place, and $\dot N^R=0$ for any $u$.  We thus conclude that the
backscattering rate $\dot n^R$ can be found for the simplest case of small
$\Delta \mu$, vanishing $u$, and unperturbed (position-independent) values
of the temperature ${\cal T} = T$ and average chemical potential $\mu$.

The resulting problem is equivalent to the one considered in
Sec.~\ref{sec:moderately_short}.  The only difference is that because the
length of the wire was assumed to be short,  $L\ll l_{\rm eq}$, the
parameter $\Delta\mu$ coincided with $eV$.  Thus replacing $eV\to \Delta
\mu$ in Eq.~(\ref{eq:NRdot_holes_result}) we find
\begin{equation}
\label{mnr}
\dot{N}^R = - r_1 \frac{2\Delta\mu}{h}.
\end{equation} 
Here the dimensionless parameter $r_1$ is defined as
\begin{align}
\label{r1}
r_1 = \frac{L}{l_1 e^{\mu/T}},
\end{align} 
and the length $l_1$ is given by Eq.~(\ref{eq:l_1_precise}).

Our expression (\ref{mnr}) and the earlier result (\ref{nri}) relate $\dot
N^R$ to two different parameters of the distribution function
(\ref{dfus}), namely, $\Delta\mu$ and $u$.  Both of these parameters
affect the electric current in the wire and can be expressed in terms of
the drift velocity $v_d=I/ne$ and the degree of equilibration $\eta$.
Indeed, according to the definition (\ref{defeta}) of $\eta$ and the
expression (\ref{currentudmu}) for the current, we have
\begin{equation}
 \label{eq:u-Delta_mu_vs_eta}
 nu=\eta\, nv_d,
\quad
 \frac{2\Delta\mu}{h}=(1-\eta)nv_d.
\end{equation}
Using these expressions to compare Eqs.~(\ref{nri}) and (\ref{mnr}), we
readily find the following expression for the parameter $\eta$
\begin{equation}
\eta = \frac{r_1}{r_0 + r_1}=\frac{L}{l_{\rm eq}+L},
\label{eq:eta_final}
\end{equation} 
where we have introduced the equilibration length 
\begin{equation}
\label{leq}
l_{\rm eq}=r_0 l_1 e^{\mu/T}.
\end{equation} 
As expected, parameter $\eta$ grows with the length of the wire from 0 to
1.  Because the backscattering process involves rare holes at the bottom
of the band, the equilibration length (\ref{leq}) at which the crossover
occurs is exponentially long at low temperatures.

Substituting Eq.~(\ref{eq:eta_final}) into (\ref{g}) we find the following
expression for the conductance of a quantum wire
\begin{equation}
 \label{eq:G_final}
 G= {2e^2\over h}
  \left[ 1- {\pi^2\over 12} \left({T\over \mu}\right)^2 
        \frac{L}{l_{\rm eq}+L}\right],
\end{equation}
valid for $L\gg l_1$.  At $L\to\infty$ it recovers the long wire limit
(\ref{eq:conductance_RMM}), while at $l_1\ll L\ll l_{\rm eq}$ it agrees
with our earlier result (\ref{g1}).

\section{Thermoelectric properties of long wires}
\label{sec4}

We now turn to a situation in which the leads are not only biased by a
finite voltage $V$ but also exposed to a temperature drop $\Delta T$. More
specifically, we assume that the leads supply the wire with the following
electron distribution
\begin{equation}
\label{distributionsdt}
f_p^{(0)} = \frac{\theta(p)}{e^{(\epsilon_p - \mu-eV)/(T+\Delta T)}+1}
          +\frac{\theta(-p)}{e^{(\epsilon_p - \mu)/T}+1}.
\end{equation}
Our goal is to find the thermopower $S$, Peltier coefficient $\Pi$, and
the thermal conductance $K$ of the quantum wire.  These transport
coefficients are defined by the following linear response relations
\begin{eqnarray}
 \label{eq:thermopower_definition}
 V&=&-S\Delta T\big|_{I=0},
\\
 \label{eq:Peltier_definition}
 j_Q&=&\Pi I\big|_{\Delta T=0},
\\
 \label{eq:thermal_conductance_definition}
 j_Q&=&K\Delta T\big|_{I=0}.
\end{eqnarray}
The thermopower and Peltier coefficient are not independent properties of
the system; they are connected by an Onsager relation $\Pi=ST$.

In the absence of interactions, electron distribution in the wire is given
by Eq.~(\ref{distributionsdt}), and its transport coefficients are easily
understood.  For instance, the thermopower and Peltier coefficient are
given by
\begin{equation}
 \label{eq:thermopower-Peltier_noninteracting}
 \Pi=ST=\frac{1}{e}[\mu e^{-\mu/T} + T(1+e^{-\mu/T})\ln(1+e^{-\mu/T})].
\end{equation}
At low temperature $T\ll\mu$, the expression
(\ref{eq:thermopower-Peltier_noninteracting}) is exponentially small,
$\Pi\sim (\mu/e)\,e^{-\mu/T}$.  The reason is that contributions to the heat
current from electrons with energies $\mu+\xi$ and $\mu-\xi$ cancel each
other, and the only reason $\Pi$ does not vanish completely is the absence of
electronic states below the bottom of the band.  In this paper we are not
interested in such exponentially small results, unless the smallness can
be compensated by a long length of the wire.  Thus to first approximation
the thermopower and Peltier coefficient of a non-interacting quantum wire
vanish.  The latter conclusion can be easily obtained from the so-called
Cutler-Mott formula\cite{mott}
\begin{equation}
 \label{eq:Mott_formula}
 S=\frac{\pi^2T}{3e} \frac{d \ln G}{d\epsilon_F},
\end{equation}
generally applicable to systems of non-interacting electrons at
$T\ll\epsilon_F$.  Considering that the conductance $G=2e^2/h$ does not
depend on the Fermi energy $\epsilon_F$, one easily concludes that $S=0$.

To the same accuracy, i.e., neglecting corrections small as $e^{-\mu/T}$, the 
thermal conductance of a non-interacting quantum wire is
\begin{equation}
 \label{eq:thermal_conductance_noninteracting}
 K=\frac{2\pi^2}{3h}\, T.
\end{equation}
This expression can be derived by straightforward calculation of the heat
current for the electron distribution (\ref{distributionsdt}) with $V=0$.
Alternatively, one can obtain
(\ref{eq:thermal_conductance_noninteracting}) from the Wiedemann-Franz law
\begin{equation}
 K=\frac{\pi^2}{3e^2}\,TG,
 \label{eq:Wiedemann-Franz}
\end{equation}
by substituting the quantized conductance $G=2e^2/h$.

It is important to note that both the Cutler-Mott formula
(\ref{eq:Mott_formula}) and Wiedemann-Franz law (\ref{eq:Wiedemann-Franz})
are not generally applicable to systems where inelastic scattering of
electrons plays an important role.  Thus one cannot expect to find the
transport coefficients $S$, $\Pi$, and $K$ in long quantum wires by
combining these relations with the expression (\ref{eq:G_final}) for the
conductance.  Below we find the transport coefficients of a long wire,
whose length $L\sim l_{\rm eq}\propto e^{\mu/T}$.  We will see that in
such wires the thermopower and Peltier coefficient are no longer
exponentially small, whereas the thermal conductance $K$ is suppressed at
$L\to\infty$.

Unlike the thermopower and thermal conductance, the Peltier coefficient
$\Pi$ is defined in a system to which no temperature bias is applied, see
Eq.~(\ref{eq:Peltier_definition}).  One can therefore obtain $\Pi$ using
the results of Sec.~\ref{sec2}.  In particular, we saw that in a long wire
the heat current is determined by the parameter $u$ of the distribution
function, see Eq.~(\ref{jq}).  The value of $u$ depends on the length of
the wire via expressions (\ref{eq:u-Delta_mu_vs_eta}) and
(\ref{eq:eta_final}).  Combining these results we find the heat current in
the form
\begin{equation}
 \label{eq:heat_current}
 j_Q=\frac{\pi^2}{6}\,
     \frac{T^2}{\mu}\,
     \frac{L}{l_{\rm eq}+L}\,
     nv_d.
\end{equation}
The ratio of $j_Q$ and the electric current $I=env_d$ gives the Peltier
coefficient 
\begin{equation}
 \label{eq:Peltier_final}
 \Pi=\frac{\pi^2}{6e}\,
     \frac{T^2}{\mu}\,
     \frac{L}{l_{\rm eq}+L}.
\end{equation}
Similar to our main result (\ref{eq:G_final}) for the conductance,
Eq.~(\ref{eq:Peltier_final}) is applicable at $L\gg l_1$.  It shows how
$\Pi$ grows from exponentially small values at $L\ll l_{\rm eq}$ to
$\pi^2T^2/6e\mu$ at $L\to\infty$.

The thermopower and thermal conductance are defined in a system with a
small temperature bias $\Delta T$.  To find these transport coefficients
we revise our analysis of the conservation laws (\ref{jl+jr}) and
(\ref{eq:heat_conservation}) to add finite $\Delta T$.  The left-hand
side of Eq.~(\ref{jl+jr}) represents the particle current $j=I/e$ in the
wire with electron distribution (\ref{distributionsdt}).  As we saw above,
the thermopower of such a wire
(\ref{eq:thermopower-Peltier_noninteracting}) is exponentially small, and
thus the effect of temperature bias on the current $j$ is negligible.  We
thus have $j=2eV/h$ and recover Eq.~(\ref{j2}).  

The left-hand side of Eq.~(\ref{eq:heat_conservation}) is the heat current
in a wire with electron distribution (\ref{distributionsdt}), which does
not vanish at $\Delta T\neq 0$.  It is determined by the thermal
conductance (\ref{eq:thermal_conductance_noninteracting}) of a
non-interacting wire.  Thus instead of Eq.~(\ref{hc}) we obtain
\begin{equation}
\label{jqc2}
\frac{2\pi^2}{3h}\, T \Delta T=j_Q-\dot{Q}^R.
\end{equation} 

The right-hand sides of equations (\ref{j2}) and (\ref{jqc2}) are not
directly related to the voltage and temperature bias of the wire, but are
determined by parameters $\Delta \mu$ and $u$ of the partially
equilibrated distribution (\ref{dfus}).  Indeed, the currents $j$ and
$j_Q$ do not depend on position, and can be calculated for the internal
region of the wire using Eqs.~(\ref{currentudmu}) and (\ref{jq}).  The
backscattering of the right-moving electrons predominantly happens inside
the wire, at distances over $l_1$ from the leads, where the partially
equilibrated distribution (\ref{dfus}) is established.  Thus we can
express $\dot N^R$ and $\dot Q^R$ in terms of $\Delta \mu$ using the
results (\ref{mnr}) and (\ref{eq:QRdot-NRdot}) of Sec.~\ref{sec2}.  As a
result, we obtain the following two linear equations upon the parameters
$\Delta\mu$ and $u$,
\begin{subequations}
 \label{eq:system-of-equations}
\begin{eqnarray}
 \label{eq:equation1}
 (1+r_1)\frac{2\Delta\mu}{h} + nu &=& \frac{2eV}{h},
\\
 -r_1\frac{2\Delta\mu}{h} + r_0nu &=& \frac{\pi^2}{3h}\, 
                                      \frac{T}{\mu}\Delta T.
\end{eqnarray}
\end{subequations}
The system of equations (\ref{eq:system-of-equations}) can be easily
solved.  Then, substituting the resulting $\Delta\mu(V,\Delta T)$ and
$u(V,\Delta T)$ into Eqs.~(\ref{currentudmu}) and (\ref{jq}), one finds
the electric current $I$ and heat current $j_Q$.

On the other hand, it is easier to find the thermopower $S$ and thermal
conductance $K$ by noticing that their definitions
(\ref{eq:thermopower_definition}) and
(\ref{eq:thermal_conductance_definition}) assume the condition of zero
current $I$.  Then, from Eq.~(\ref{currentudmu}) we immediately find
$  \frac{2\Delta\mu}{h}=-nu$ and equations (\ref{eq:system-of-equations})
reduce to
\begin{subequations}
 \label{eq:system-of-equations_reduced}
\begin{eqnarray}
 \label{eq:equation1_reduced}
 -r_1 nu &=& \frac{2eV}{h},
\\
 \label{eq:equation2_reduced}
 (r_1 + r_0)nu &=& \frac{\pi^2}{3h}\, \frac{T}{\mu}\Delta T.
\end{eqnarray}
\end{subequations}
Excluding the unknown parameter $u$, we find the linear relation
(\ref{eq:thermopower_definition}) between $V$ and $\Delta T$ with
\begin{equation}
 \label{eq:thermopower_final}
 S=\frac{\pi^2}{6e}\,
     \frac{T}{\mu}\,
     \frac{L}{l_{\rm eq}+L}.
\end{equation}
Predictably, the thermopower (\ref{eq:thermopower_final}) and the Peltier
coefficient (\ref{eq:Peltier_final}) satisfy the Onsager relation
$\Pi=ST$.

Furthermore, using Eq.~(\ref{eq:equation2_reduced}) we express the heat
current (\ref{jq}) in terms of $\Delta T$ as 
\begin{equation}
 \label{eq:heat-corrent_vs_DeltaT}
 j_Q=\frac{2\pi^2T}{3h}\, \frac{r_0}{r_1+r_0}\Delta T.
\end{equation}
Thus the thermal conductance takes the form
\begin{equation}
 \label{eq:thermal_conductance_final}
 K=\frac{2\pi^2T}{3h}\, \frac{l_{\rm eq}}{l_{\rm eq}+L}.
\end{equation}
At $L\ll l_{\rm eq}$, Eq.~(\ref{eq:thermal_conductance_final}) recovers the
result (\ref{eq:thermal_conductance_noninteracting}) for noninteracting
wires, but as the length of the wire grows, $K$ is suppressed as $1/L$.

The fact that the thermal conductance $K$ vanishes at $L\to\infty$ can be
understood as follows.  In an infinitely long wire the distribution
function of electrons reaches the fully equilibrated form
(\ref{vddistrib}) controlled by three parameters, $T$, $\mu$, and the
drift velocity $v_d$.  The thermal conductance is defined under the
condition that the electric current $I=env_d$ vanishes, see
Eq.~(\ref{eq:thermal_conductance_definition}).  Thus $v_d=0$ and the
distribution (\ref{vddistrib}) takes the form of the standard Fermi-Dirac
distribution.  Due to its symmetry $p\to-p$, the heat current $j_Q$
vanishes, regardless of the temperature bias $\Delta T$ applied to the
wire.  One therefore finds that in an infinitely long wire $K=0$.

The thermoelectric properties of a device are sometimes summarized in the
form of the dimensionless figure of merit defined as
\begin{align}
Z T = {G S^2 T\over K}.
\end{align} 
The figure of merit measures the efficiency of thermoelectric
refrigerators.  As $ZT$ diverges, the device attains Carnot efficiency.
For a material to be a good thermoelectric cooler, it must have a high
value for $ZT$ and a typical figure of merit $ZT \simeq 3$ would make
solid-state home refrigerators economically competitive with
compressor-based refrigerators.\cite{mahan}  However, in many materials the
figure of merit is limited by the Wiedemann-Franz law, and remains near 1.

Substituting our results (\ref{eq:thermopower_final}) and
(\ref{eq:thermal_conductance_final}) for a long wire and neglecting a
small correction to the quantized conductance, we find
\begin{equation}
\label{ztl}
Z T \simeq \frac{\pi^2}{12} 
            \left( \frac{T}{\mu} \right)^2 
            \frac{L^2}{l_{\rm eq}(l_{\rm eq}+L)}.
\end{equation} 
In short wires the thermopower is small, resulting in a small $ZT$.  On
the other hand, the thermal conductance $K$ is strongly suppressed in long
wires, giving rise to infinite figure of merit at $L\to\infty$.

\section{Discussion}
\label{sec5}

\begin{figure}[tb]
\centering
\resizebox{.45\textwidth}{!}{\includegraphics{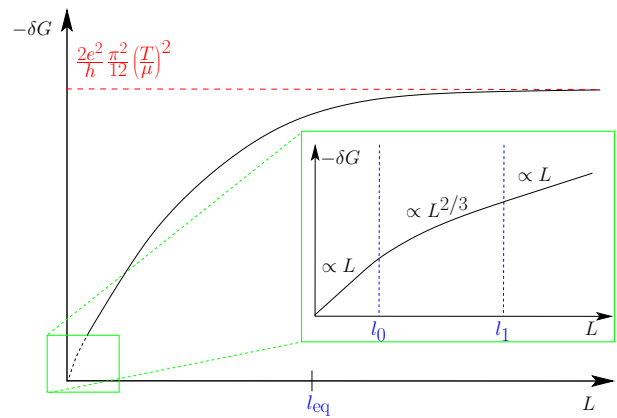}}
\caption{Correction to the conductance of a quantum wire as a function of
 its length.  For wires longer than the scale $l_1$ the conductance is
 given by Eq.~(\ref{eq:G_final}).  The behavior of the conductance in the
 regions $L\ll l_0$ and $l_0\ll L\ll l_1$ is given by
 Eqs.~(\ref{eq:deltaG_Lunde}) and (\ref{g2}), respectively.}
\label{summary}
\end{figure}

In this paper we studied the transport properties of a partially
equilibrated quantum wire.  In one-dimensional systems, equilibration of
weakly interacting electrons is strongly suppressed at low temperatures,
and the resulting equilibration length $l_{\rm eq}$ is exponentially
large, Eq.~(\ref{leq}).  Our main result is the expression
(\ref{eq:G_final}) for the conductance of a wire whose length $L$ exceeds
the length scale $l_1$ given by Eq.~(\ref{eq:l_1_precise}).  Because the
scale $l_1$ is only power-law large at low temperature, the expression
(\ref{eq:G_final}) describes the full crossover behavior of conductance
between the regimes of negligible and full equilibration.  We have also
been able to establish a connection between our result (\ref{eq:G_final})
and the expression (\ref{eq:deltaG_Lunde}) for the correction to the
conductance of a short wire obtained by Lunde \emph{et al.}\cite{lunde1}.
Similar to Eq.~(\ref{eq:deltaG_Lunde}), our result (\ref{eq:G_final}) is
exponentially suppressed at small $L$ and grows linearly with $L$.
However the prefactors are parametrically different.  This mismatch is
resolved by noticing that Eq.~(\ref{eq:deltaG_Lunde}) is valid at $L\ll
l_0$, where the length $l_0$ defined by Eq.~(\ref{eq:mean_free_path}) is
short compared to $l_1$.  In the regime of intermediate wire lengths,
$l_0\ll L \ll l_1$, the correction to the conductance (\ref{g2}) scales
with the length as $L^{2/3}$.  A summary of our results for the
conductance of a quantum wire as a function of its length is presented in
Fig. \ref{summary}.

In addition to conductance, we studied thermoelectric effects in partially
equilibrated wires, limiting ourselves to the most interesting regime
$L\gg l_1$.  The equilibration of the electron system has a dramatic
effect on the thermopower and thermal conductance of the wire.  As the
length of the wire increases, the thermopower increases dramatically, from
exponentially small values at $L\ll l_{\rm eq}$ to $S\sim T/e\mu$ at $L\gg
l_{\rm eq}$, see Eq.~(\ref{eq:thermopower_final}).  Conversely, the
thermal conductance of the wire decreases due to the equilibration of the
electron system in the wire from the Wiedemann-Franz value $K=2\pi^2 T/3h$
to zero, Eq.~(\ref{eq:thermal_conductance_final}).  As a result, at $L\gg
l_{\rm eq}$ the quantum wire becomes a perfect thermoelectric refrigerator.

In this paper we accounted for the effect of electron-electron
interactions but neglected the electron-phonon scattering, which may also
affect the electron distribution function.  In GaAs quantum wires the
phonons are three-dimensional and in equilibrium with the rest of the
system.  Scattering of electrons by phonons should therefore have the
effect of equilibrating them in the stationary reference frame, thereby
reducing the degree of equilibration $\eta$.  We leave the detailed study
of the effect of phonon coupling to future work and limit our discussion
here to a few qualitative remarks.  The reason the electron-phonon
coupling is typically neglected compared to electron-electron interactions
is that respective coupling constant is much smaller for the phonons.  On
the other hand, we do not expect the effect of phonons on the electron
distribution function to be exponentially suppressed as $e^{-\mu/T}$.
Thus we expect that coupling to the phonons to be negligible only at not
too low temperatures. 

Another effect neglected in this paper is the possible presence of slight
long-range inhomogeneities in the wire, which would typically be caused by
the presence of remote impurities in the GaAs heterostructure.  The effect
of such inhomogeneities on the conductance was studied
earlier\cite{jerome1,jerome2} under the assumption of full equilibration
of the electron system.  On the other hand, the inhomogeneities themselves
resist the equilibration process, and we expect an interesting interplay of
these effects in the case of a partially equilibrated wire.  We leave such
a study to future work.

\subsection*{Acknowledgements}

We are grateful to B.~L. Altshuler, A.~V. Andreev, and L.~I. Glazman for
helpful discussions.  This work was supported by the U.S. Department of
Energy, Office of Science, under Contract No. DE-AC02-06CH11357, and
through SFB/TR 12 of the Deutsche Forschungsgemeinschaft (DFG), the Center
for Nanoscience (CeNS) Munich, and the German Excellence Initiative via
the Nanosystems Initiative Munich (NIM).

\begin{appendix}

\section{Screened Coulomb interaction}
\label{aem}

Let us consider the Coulomb interaction between electrons screened by a nearby gate, which we model by a conducting plane at a distance $d$ from the wire. In this case, the electron-electron interaction takes the form
\begin{align}
\label{real}
V(x)= {e^2\over \epsilon}
\left( {1\over |x|} -{1\over\sqrt{x^2+(2d)^2}}\right).
\end{align}
The diverging short-range behavior of this potential needs to be regularized in order to evaluate the small-momentum Fourier components $V_q$. To this end, we introduce the small width $w$ of the quantum wire, $w \ll d$. Then the homogeneous component $V_0$ of the interaction potential takes the form
\begin{align}
\label{hom}
V_0 = {2e^2\over \epsilon}  \ln \left( \frac{d}{w} \right).
\end{align} 

For small wave vectors $q \ll 1/d$, the Fourier transformed potential $V_q$ departs from the homogeneous component $V_0$ by an amount
\begin{align}
V_q - V_0 &= {e^2\over \epsilon} \int dx \left[\cos( q x ) - 1\right]
\left( {1\over |x|} -{1\over\sqrt{x^2+(2d)^2}}\right) \nonumber \\
& \simeq - {2e^2\over \epsilon} d^2 q^2 \ln \left( \frac{1}{|q| d} \right) .
\end{align} 
It then follows that the small-$q$ behavior of the Fourier-transformed potential is given by
\begin{align}
V_q = V_0 \left[ 1 - q^2 d^2 \frac{\lnæ( 1 / |q| d)}{\ln ( d /wæ)} \right] .
\label{final_vq}
\end{align}
This expression contains an extra logarithmic-in-$q$ factor compared to the expression introduced by Lunde {\it et al.},\cite{lunde1}
\begin{align}
\label{vqlunde}
V_q = V_0 \left[ 1 - \frac{q^2}{q_0^2} \right] .
\end{align}
However, as argued in the text, the typical scattering processes studied here only involve small momentum exchanges, of the order $\hbar q \sim T/{v_F}$, so that the expression (\ref{final_vq}) for $V_q$ reduces to the one of Eq.~(\ref{vqlunde}) with
\begin{align}
\label{redef_q0}
q_0 = \frac{1}{d} \left( \frac{\lnæ( \hbar v_F / T d)}{\ln ( d /wæ)}  \right)^{1/2} .
\end{align}
The model introduced in Eq.~(\ref{real}), therefore merely amounts to an extra logarithmic temperature dependence in the length $l_{eee}$, Eq.~(\ref{lunde}).

\section{Fokker-Planck equation for 3-particle collisions}
\label{afp}
In this appendix we discuss the Fokker-Planck 
approximation and calculate the coefficients for the interaction potential
used by Lunde {\it et al.} \cite{lunde1} as well as the screened and 
unscreened Coulomb interaction.

\subsection{Fokker-Planck approximation}

We start out from the collision integral (\ref{collisionintegral})
for the three-particle scattering process.
We discussed in section~\ref{sec1} that the only contributions 
relevant to transport result from collisions 
involving two pairs of incoming and outgoing 
states in the vicinity of the right and left Fermi-points, and one pair 
of incoming and outgoing states at the bottom of 
the band, see Fig.~\ref{fig2}b. Let $p_1$ and $p'_1$ be the momenta 
near the bottom of the band, $p_2$ and 
$p'_2$ the ones near the left Fermi point, while $p_3$ and $p'_3$ are taken
near the right Fermi point.
Unprimed momenta correspond to incoming states 
whereas primed ones are associated with outgoing states.
We  introduce the hole distribution 
$g_{p_i}=1-f_{p_i}$ and the collision integral of holes $I_{p,x}[g] 
= - I_{p,x}[f] $, which using Eq.~(\ref{collisionintegral}),  can be recast as
\begin{align}
\label{app12}
I_{p_1,x}[g] 
&= \sum_{p'_1} \left[ W(p_1,p'_1)g_{p'_1} - W(p'_1,p_1)g_{p_1} \right],
\end{align} where 
\begin{align}
\label{app13}
W(p_1,p'_1) 
&=  48 \sum_{\substack{p_2,p_3\\ p'_2,p'_3}} W_{123;1'2'3'}
g_{2'}g_{3'}f_1f_2f_3, \\
W(p'_1,p_1)
&= 48 \sum_{\substack{p_2,p_3\\ p'_2,p'_3}} W_{123;1'2'3'} g_2g_3 f_{1'}f_{2'}f_{3'}.
\label{app13-2}
\end{align} 
$W(p_1,p'_1)$ is the rate for a transition in which a hole scatters from
some state $p'_1$ into $p_1$, while $W(p'_1,p_1)$ denotes the
corresponding transition rate for the inverse process.  Here and in what
follows, all momentum summations are restricted to the ranges discussed
above.  This restriction results in a combinatorial factor of 12 in
Eqs.~(\ref{app13}) and (\ref{app13-2}).  The remaining factor of 4
originates from the spin summations as we anticipated that the main
contribution to the 3-particle scattering rate $w_{123;1'2'3'}$ of
Eq.~(\ref{collisionintegral}) takes the form $\delta_{\sigma_1
 \sigma_{1'}} \delta_{\sigma_2 \sigma_{2'}} \delta_{\sigma_3 \sigma_{3'}}
W_{123;1'2'3'}$, with a spin-independent $W_{123;1'2'3'}$. This
simplification is only valid in the limit of small momentum exchanges, and
can be performed here since for the Coulomb interaction $V_0 \gg V_{k_F}$.
Since $p_1$ and $p'_1$ lie near the bottom of the band, the distribution
functions $g_{p_1}$ and $g_{p'_1}$ are exponentially small, and so is the
collision integral of holes (\ref{app12}). It is therefore unnecessary to
account for additional exponentially small contributions in the scattering
rates $W(p_1,p'_1)$ and $W(p'_1,p_1)$, so that one can safely replace $f_1
\simeq 1$ and $f_{1'} \simeq 1$ in Eqs.~(\ref{app13}) and (\ref{app13-2}).

The Fokker-Planck approximation exploits the fact that collisions
typically induce small momentum changes of order ${\cal O}(T/v_F)$.  For
the following, it is convenient to introduce the momentum exchanges
$q_1=p'_1-p_1$.  With this notation, $W(p'_1,p_1)$ describes the
transition rate for the process in which a hole scatters with momentum
transfer $q_1$, from the initial state $p_1$, and can thus be rewritten as
$W(p'_1,p_1)= W_{q_1}(p_1)$. Following the same prescription, the
transition rate for the inverse process becomes $W(p_1,p'_1) =
W_{-q_1}(p_1+q_1)$. Performing a small-momentum expansion, one has
\begin{align}
\label{qepxansion}
W_{-q_1} & (p_1+q_1) g_{p'_1} 
=  W_{-q_1} (p_1) g_{p_1} + q_1 
\partial_{p_1} \left(W_{-q_1}(p_1) g_{p_1} \right)  \nonumber \\ 
& 
+ {q_1^2\over 2}  \partial^2_{p_1} \left(W_{-q_1}(p_1) g_{p_1} \right) 
+ {\cal O}\left( q_1^3 \partial^3_{p_1} (W g) \right),
\end{align} where $\partial_{p_i}=\partial/\partial p_i$.
Introducing further
\begin{align}
\label{app14}
A(p_1)=& - \sum_{q_1} q_1 W_{-q_1}(p_1) &= \sum_{q_1} q_1 W_{q_1}(p_1) , \\
\label{app14-2}
B(p_1)=& \sum_{q_1} q^2_1 W_{-q_1}(p_1)&= \sum_{q_1} q^2_1 W_{q_1}(p_1), 
\end{align} the collision integral of holes takes the simplified form 
\begin{align}
\label{app15}
I_{p_1,x}[g] =
&  - \partial_{p_1} \left( A(p_1) g(p_1) \right) 
+ {1\over 2} \partial_{p_1}^2 \left( B(p_1) g(p_1) \right) .
\end{align} 
We next turn to the explicit derivation of the functions
$A(p)$ and $B(p)$ in the case of three-particle collisions.

\subsection{Relation between $A(p)$ and $B(p)$}

The scattering rate $W_{123;1'2'3'}$ contains both the energy and momentum conservation and can be rewritten as
\begin{align}
W_{123;1'2'3'} =& \delta (\epsilon_{1'}-\epsilon_1+\epsilon_{2'}-\epsilon_2+\epsilon_{3'}-\epsilon_3) \nonumber \\
& \times \delta_{q_1+q_2+q_3,0}  ~w (q_1,  q_2,  q_3) ,
\label{explicit_W}
\end{align}
where we introduced the momentum transfers $q_i = p'_i-p_i$. The function
$w$ that remains after writing the conservation laws explicitly, should
depend on all $p_i$ and $p_{i'}$. However, for the momentum configuration
under consideration, $p_1$ lies near the bottom of the band, while $p_2$
and $p_3$ lie near the left and right Fermi points, all within a small
range set by temperature. We thus argue that, up to small corrections in
$T/\mu$, one can replace $p_1 \simeq 0$, $p_2 \simeq - p_F$ and $p_3
\simeq p_F$ in the expression for $w$, which then becomes a function of
$q_1$, $q_2$ and $q_3$.

Using the approximated forms $\epsilon_{2'}-\epsilon_2 \simeq - v_F q_2$
and $\epsilon_{3'}-\epsilon_3 \simeq v_F q_3$, the conservation laws allow
us to express $q_2$ and $q_3$ in terms of $p_1$ and $p'_1$ as
\begin{align}
\label{app18}
q_2 &= \frac{p_1-p'_1}{2} + \frac{\epsilon_{1'}-\epsilon_1}{2 v_F}  ,\\
q_3 &= \frac{p_1-p'_1}{2} - \frac{\epsilon_{1'}-\epsilon_1}{2 v_F}  ,
\label{app18-2}
\end{align}
where one readily sees that $q_2 \simeq q_3 \simeq -q_1 /2$, up to small
contributions of order $p_1/p_F \ll 1$.

Substituting the expression (\ref{explicit_W}) for the scattering rate
into Eq.~(\ref{app13-2}), and using the energy and momentum conservation
laws to simplify two of the momentum summations, one has
\begin{align}
\label{app16}
W(p'_1, p_1) = 48  \frac{\Delta L}{h v_F}  w(q_1, q_2, q_3) 
              \sum_{p_2,p_3}  g_{p_2} g_{p_3} f_{p_2+q_2}f_{p_3+q_3},
\end{align} 
where we focused on a section of the wire, of length $\Delta L \ll l_{\rm
 eq}$. Here $q_2$ and $q_3$ are functions of $p_1$ and $p'_1$, as given
by Eqs.~(\ref{app18}) and (\ref{app18-2}).

The remaining momentum summations can be performed explicitly upon
linearizing the dispersion near the Fermi level
\begin{align}
\label{app17}
\sum_{p_2} g_{p_2} f_{p_2+q_2} &= - \frac{\Delta L}{h} \frac{q_2 e^{\beta v_F q_2}}{1-e^{\beta v_F q_2}}, \\
\sum_{p_3} g_{p_3} f_{p_3+q_3} &= - \frac{\Delta L}{h} \frac{q_3}{1-e^{\beta v_F q_3}},
\label{app17-2}
\end{align}
so that the transition rate, Eq.~(\ref{app16}) becomes
\begin{align}
\label{app19}
W(p'_1,p_1) = \frac{3}{v_F} \left( \frac{\Delta L}{h} \right)^3 \frac{q_1^2 ~w_{q_1}}{\sinh^2 \left( \frac{\beta v_F q_1}{4}\right)} \left( 1 + \frac{\epsilon_{1'}-\epsilon_1}{2 T} \right),
\end{align}
where we replaced $q_2$ and $q_3$ following Eqs.~(\ref{app18}) and
(\ref{app18-2}), and introduced $w_{q_1} = w(q_1, -q_1/2, -q_1/2)$. The
leading contribution to $W(p'_1,p_1)$ is an even function of $q_1$ which
leads to a vanishing $A(p_1)$ once substituted into Eq.~(\ref{app14}). For
that reason, we expanded the expression for the transition rate up to
linear order in the small parameter $(\epsilon_{1'}-\epsilon_1) / T\sim
p_1/p_F \ll 1$.

The functions $A(p_1)$ and $B(p_1)$ are then readily obtained from
Eqs.~(\ref{app14}) and (\ref{app14-2}) by substituting the expression
(\ref{app19}) above for $W(p'_1,p_1)$ yielding
\begin{align}
A(p_1) &= \frac{1}{2} \sum_{q_1} q_1 \left[ W(p_1+q_1,p_1) - W(p_1-q_1,p_1) \right] \nonumber \\
&= \frac{p_1}{2 m T} \sum_{q_1} \frac{3}{v_F} \left( \frac{\Delta L}{h} \right)^3 \frac{q_1^4 ~w_{q_1}}{\sinh^2 \left( \frac{\beta v_F q_1}{4}\right)}, 
\label{apa}
\end{align}
and
\begin{align}
\label{apb}
B(p_1) &= \frac{1}{2} \sum_{q_1} q_1^2 \left[ W(p_1+q_1,p_1) + W(p_1-q_1,p_1) \right] \nonumber \\
&= \sum_{q_1} \frac{3}{v_F} \left( \frac{\Delta L}{h} \right)^3 \frac{q_1^4 ~w_{q_1}}{\sinh^2 \left( \frac{\beta v_F q_1}{4}\right)},
\end{align} 
where we discarded contributions of order $(p_1/p_F)^2$ and higher.  
It follows from these two expressions that for a 
momentum $p_1$ deep in the band, $|p_1| \ll p_F$, 
the function $B(p_1)$ can be approximated by a 
constant $B$, while $A(p_1)$ satisfies
\begin{align} 
A(p_1)= {p_1\over 2mT} B.
\end{align}
In order to derive an explicit form of the Fokker-Planck equation, it is
therefore sufficient to calculate the constant $B$.

Let us now briefly comment on the validity of the Fokker-Planck
approximation.  The first two terms neglected within the Fokker-Planck
approximation would contribute to the collision integral as
$\partial_{p_1}^3 (C(p_1) g_{p_1})$ and $\partial_{p_1}^4 (D(p_1)
g_{p_1})$, where $C(p_1) = \sum_{q_1} q_1^3 W_{q_1}(p_1)$ and $D(p_1) =
\sum_{q_1} q_1^4 W_{q_1}(p_1)$.  Going through the same derivation as the
one outlined above, and keeping in mind that every new power of $q$
results in a factor of $T/v_F$, one can convince oneself that
$C(p_1)\propto \left({T\over v_F}\right)^2 A(p_1)$ and $D(p_1) \propto
\left({T\over v_F}\right)^2 B$.  It results that the contribution to the
collision integral from the terms in $C(p_1)$ and $D(p_1)$ are smaller than
the ones from $A(p_1)$ and $B(p_1)$ by a factor $T/\mu \ll 1$.  This
readily generalizes to higher order derivatives $\partial_p^n$, thus
validating the expansion of the collision integral used here.

\subsection{Evaluation of $B$}

We now derive the expression for the constant $B$ using 
the specific form of the electron-electron interaction potential of Eq.~(\ref{vqlunde}).
This expression of the potential is 
largest for small wave vectors $q$ 
allowing to discard the exchange terms  
$|V_{|q|\sim k_F} | \ll |V_{|q|\ll k_F}|$ in the scattering rate, which  is thus 
dominated by the direct term.  Following Ref.~[\onlinecite{lunde1}], the reduced scattering rate 
$w_q$ takes the form
\begin{align}
\label{avq}
w_q
= \frac{2\pi}{\hbar} 
\left[ {V_{q/2}\left(V_{q}-V_{q/2}\right)\over 2\Delta L^2\mu}\right]^2 .
\end{align}
Expanding for small values of $q$, this can be further rewritten as
\begin{align}
\label{applunde}
w_q 
= \frac{9 \pi^2}{16 h} \frac{1}{(k_F \Delta L)^4} 
\frac{(V_0 k_F)^4}{\mu^2} \left( \frac{q}{q_0} \right)^4,
\end{align} 
where $V_0$ is the zero-momentum Fourier component of the potential. 
Substituting this expression back into 
Eq.~(\ref{apb}), and performing the integral over $q_1$, 
we finally find for the constant $B$
\begin{align}
\label{appb0}
B = \frac{9 \pi^5}{20} 
\left( \frac{V_0 k_F}{\mu} \right)^4 
\left( \frac{k_F}{q_0} \right)^4 
\left( \frac{T}{\mu} \right)^9  k_Fp_F \mu,
\end{align}
from which we can extract the length scales $l_1$, $l_{eee}$ and $l_0$ using Eqs.(\ref{eq:l_1_precise}), (\ref{eq:l_1}) and (\ref{eq:mean_free_path}) respectively
\begin{align}
l_1^{-1} & = \frac{9 \pi^{9/2}}{40} \left( \frac{V_0 k_F}{\mu} \right)^4 
\left( \frac{k_F}{q_0} \right)^4 
\left( \frac{T}{\mu} \right)^{15/2} k_F , \\
l_{eee}^{-1} & \sim \left( \frac{V_0 k_F}{\mu} \right)^4 
\left( \frac{k_F}{q_0} \right)^4 
\left( \frac{T}{\mu} \right)^7 k_F , \\
l_0^{-1} & \sim \left( \frac{V_0 k_F}{\mu} \right)^4 
\left( \frac{k_F}{q_0} \right)^4 
\left( \frac{T}{\mu} \right)^6 k_F .
\label{lunde_l0}
\end{align}
The expression for $B$ is model-specific, and the result of
Eq.~(\ref{appb0}) was obtained for the potential $V_q$ of
Eq.~(\ref{vqlunde}), leading to the same expression for $l_{eee}$ as the
one of Lunde {\it et al.},\cite{lunde1} Eq.~(\ref{lunde}). In the case of
the screened Coulomb potential discussed in Appendix~\ref{aem} and for
temperatures $T \ll \hbar v_F / d$, one obtains similar results upon
redefining $q_0$ according to Eq.~(\ref{redef_q0}). On the other hand, for
temperatures $T \gg \hbar v_F/d$, the effect of the screening gate can be
neglected, and the electron-electron interaction is then well described by
an unscreened Coulomb potential, of the form $V_q = \frac{2 e^2}{\epsilon}
\ln \left( 1/|q|w\right)$ at small wave vectors $|q| \ll w^{-1}$. This in
turn leads to the following value of $B$
\begin{align}
\label{unscr_B}
B = \frac{8 \pi \ln^2 2}{5} \left( \frac{e^2}{\epsilon} \frac{k_F}{\mu} \right)^4 \left( \frac{T}{\mu} \right)^5 \ln^2 \left( \frac{\hbar v_F}{T w} \right) k_F p_F \mu
\end{align}
and the corresponding expressions for the length scales
\begin{align}
l_1^{-1} & =   \frac{4 \sqrt{\pi} \ln^2 2}{5} \left(  \frac{e^2}{\epsilon} \frac{k_F}{\mu} \right)^4 
\left( \frac{T}{\mu} \right)^{7/2} \ln^2 \left( \frac{\hbar v_F}{T w} \right) k_F , \\
l_{eee}^{-1} & \sim \left( \frac{e^2}{\epsilon} \frac{k_F}{\mu} \right)^4 
\left( \frac{T}{\mu} \right)^3 \ln^2 \left( \frac{\hbar v_F}{T w} \right) k_F , \\
l_0^{-1} & \sim \left(  \frac{e^2}{\epsilon}\frac{k_F}{\mu} \right)^4 
\left( \frac{T}{\mu} \right)^2 \ln^2 \left( \frac{\hbar v_F}{T w} \right) k_F .
\label{unscr_l0}
\end{align}
Substituting the expression (\ref{hom}) for $V_0$, and (\ref{redef_q0}) for $q_0$ into Eq.~(\ref{appb0}), one readily recovers that Eqs.~(\ref{appb0})-(\ref{lunde_l0}) match with Eqs.~(\ref{unscr_B})-(\ref{unscr_l0}) at the crossover temperature $T \sim \hbar v_F / d$.

\section{Position dependence of the distribution function}
\label{amdf}

In this appendix we determine the profile of the position-dependent parameters 
$\mu^{R/L}(x)$ and ${\cal T}(x)$ entering the distribution function (\ref{dfus}).

As discussed in the text, steady state parameters $u$ 
and $\Delta \mu$ are constant along the wire. Therefore,  the 
spatial profile of distribution~(\ref{dfus}) is determined by space-dependencies of 
the average chemical potential and temperature. 
It is convenient to measure deviations from the lead values, 
\begin{align}
\label{parvar}
\mu^R(x) &= \mu_l + \delta\mu^R(x),  \nonumber\\
{\cal T}(x) &= T + \delta\tau(x), 
\end{align} and $\mu^L(x)=\mu^R(x)-\Delta \mu$.
Let us then consider a wire of length $L\gg l_1$, and focus on a small segment between the positions $x$
and $x+\Delta x$, where $0<x<L$. We observe that conservation of momentum insures homogeneity
of the momentum current
\begin{align}
\label{appjp}
j_P = j_P^0 
+ {hn\over 2}\left( eV + \delta\mu^R +\delta\mu^L \right) 
+ {\pi^2\over 3}{T\over \mu}p_F\delta\tau,
\end{align} 
where $j_P^0$ is the momentum current in absence of external potential bias. 
From Eq.~(\ref{appjp}) one readily observes that for 
$j_P$ to remain constant, a drop in chemical potentials must be compensated for by an increase in temperature, 
\begin{align}
\label{full_tau}
{\delta\tau(x+\Delta x)-\delta\tau(x)
\over \delta\mu^R(x+\Delta x) - \delta\mu^R(x)} 
= -{6\over \pi^2}{\mu\over T} ,
\end{align} 
valid up to small corrections in $(T/\mu)^2$. To calculate the spatial 
profile of $\mu^R$ and ${\cal T}$ we need to find the slope of either one 
and their boundary values near the leads.

The slope of $\mu^R$ is readily found from calculating 
the difference 
in right-moving particle currents within the segment of length $\Delta x$, 
\begin{align} 
\label{jR_muR}
j^R(x+\Delta x) - j^R(x)={2\over h} \left(\delta\mu^R(x+\Delta x) - \delta\mu^R(x)\right) .
\end{align} 
Since this difference equals the rate of change of the number of
right-moving electrons within the segment $(x, x+\Delta x)$, we can insert
Eq.~(\ref{mnr}) into the left hand side, to find
\begin{align} 
\label{slope_mu_R}
\delta\mu^R(x+\Delta x) - \delta\mu^R(x) = {h\dot{n}^R\over 2} \Delta x,
\end{align}
where $\dot{n}^R=\dot{N}^R/L$.  Then,
from Eq.~(\ref{mnr})   
\begin{align}
\label{dslope_mu_R}
\frac{d \mu^R(x)}{dx} = - {\pi^2\over 12}\left({T\over \mu}\right)^2 \frac{ eV}{L+l_{\rm eq}},
\end{align} 
i.e., the chemical potentials linearly decrease, while the
temperature linearly increases along the wire,
\begin{align}
\mu^R(x)&=\mu^R(0)-{\pi^2\over 12}\left({T\over \mu}\right)^2{ eV\, x\over L+l_{\rm eq}} ,\\
\mu^L(x)&=\mu^L(L)+{\pi^2\over 12}\left({T\over \mu}\right)^2  {eV(L-x)\over L+l_{\rm eq}},\\
{\cal T}(x)&={\cal T}(0)+{T\over 2\mu}{eV\,x\over L+l_{\rm eq}}.
\label{T-of-x}
\end{align}

In the linear response regime boundary values $\mu^{R}(0)$ and $\mu^{L}(L)$ 
deviate from the chemical potentials in the leads 
by an amount proportional to the applied voltage. From inversion symmetry it 
further follows that these deviations are opposite in sign, i.e.
\begin{align}
\label{app:lambdar}
\mu^R(0)&=\mu_l - \lambda eV, \\
\label{app:lambdal}
\mu^L(L)&=\mu_r + \lambda eV.
\end{align} The parameter $\lambda$ 
may be inferred from the equation
\begin{align}
\mu^R(0)- \mu^L(L) &= \mu^R(0)- \mu^R(L) -\Delta\mu,
\end{align} 
by inserting Eqs.~(\ref{app:lambdar}) and (\ref{app:lambdal}) into the
left hand side and rewriting the right hand side using
Eqs.~(\ref{dslope_mu_R}), (\ref{eq:u-Delta_mu_vs_eta}) and
(\ref{eq:G_final}). This results in
\begin{align}
\lambda = \frac{hnu}{4},
\end{align} 
where $n$ is the electron 
density.

The boundary values ${\cal T}(0)$ and ${\cal T}(L)$, are found in a
similar way by combining Eq.~(\ref{T-of-x}) with the observation that
$\delta\tau(0)=-\delta\tau(L)$, which is, again, a consequence of
inversion symmetry.  We summarize the values for the position-dependent
parameters close to the leads in terms of $eV$ and $u$
\begin{align}
\label{abce}
&\mu^R(0) = \mu + eV - {hnu\over 4},  
& \mu^L(L) =\mu+ {hnu\over 4}, \nonumber\\ 
&{\cal T}(0) = T \left(1 - {u\over v_F}\right),   
&{\cal T}(L) = T \left(1 + {u\over v_F}\right) ,
\end{align} 
where $v_F=\sqrt{2\mu/m}$. 

Finally, restricting ourselves to linear terms in $V$ and 
finite-temperature corrections to leading order 
in $(T/\mu)^2$, we observe that the values given in (\ref{abce}) 
guarantee that all moments $\langle v_p p^s \rangle$, 
with $s\in \mathbb{N}$, are continuous at the boundary 
between the wire and the leads, i.e. 
\begin{align}
\label{abc1}
\int_0^\infty {dp\over h}\, v_p p^s \left[ f_p(0)-f_p^{(0)}\right] &= 0,\\
\label{abc2}
\int_{-\infty}^0{dp\over h}\, v_p p^s \left[ f_p(L)-f_p^{(0)}\right] &= 0,
\end{align} 
where $f_p^{(0)}$ is the lead distribution function (\ref{distributions}). 
For $s =0$, 1 and 2, relations (\ref{abc1}) and (\ref{abc2}) imply continuity of particle, 
momentum and energy currents at the boundary between wire and leads. 

To show the validity of Eqs.~(\ref{abc1}) and (\ref{abc2}) we
express the distribution function (\ref{dfus}) in terms of 
$\delta\mu^{R,L}$ and $\delta\tau$, and  
expand the difference of distributions entering 
Eqs.~(\ref{abc1}) and (\ref{abc2}) to linear order 
in these parameters. For right-moving electrons close to the left lead, one has
\begin{align}
f_p(0)-f_p^{(0)} 
= -\left[ \delta\mu^R(0)+up 
+ (\epsilon_p-\mu){\delta\tau(0)\over T}\right] 
{d f_p^{(0)}\over d\epsilon},
\end{align} 
and similarly for left-movers
close to the right lead. 
Upon introducing the new variables $\xi=p^2/2m-\mu$ and $z=\xi/T$, 
and neglecting exponentially small contributions 
$\propto e^{-\mu/T}$ this results in
\begin{widetext}
\begin{align}
\int_0^\infty {dp\over h}\, v_p p^s \left[ f_p(0)-f_p^{(0)}\right] 
= - {(2m\mu)^{s \over2}\over h} \int_{-\infty}^\infty dz 
\left(1 + {zT\over \mu}\right)^{s\over2}
\left[ \delta\mu^R(0)+u\sqrt{2m\mu} 
\sqrt{1+{zT\over\mu}} + \delta\tau(0) z \right] f_0'(z),
\end{align}
where $f_0(z)=(1+e^z)^{-1}$. Keeping now only terms up to quadratic order in 
$(T/\mu)$ one then finds 
\begin{align}
&\int_0^\infty {dp\over h}\, v_p p^s\left[f_p(0)-f_p^{(0)}\right] 
= {(2m\mu)^{s \over2}\over h}
\left( \delta\mu^R(0)\left[ 1 + s(s-2){\pi^2\,T^2\over 24\,\mu^2}\right] 
+ u\sqrt{2m\mu}\left[ 1 + (s^2-1){\pi^2\,T^2\over 24\,\mu^2}\right]
+ \delta\tau(0) s {\pi^2\, T\over 6\,\mu} \right).
\label{mom_t_mu}
\end{align}
\end{widetext}  Substituting values for $\delta\mu^R(0)$
and $\delta\tau(0)$ from (\ref{abce}), one can readily check that  
Eq.~(\ref{abc1}) is satisfied for all values of $s\in \mathbb{N}$.
  Proceeding in an analogous way at the right end of the 
wire confirms Eq.~(\ref{abc2}).

\section{Energy transferred in a backscattering process}
\label{aernr}

In this appendix we calculate the change in right-movers
energy associated with the backscattering of a right-moving electron.

Let us focus on a small segment of wire in between positions $x$ and $x+\Delta x$. Following Eq.~(\ref{conteq}), we use the conservation of the number of particles to express the rate of change in the number of right-movers $\dot{N}^R$ in terms of particle currents. Proceeding similarly with the energy currents, one can express the ratio $\dot{E}^R / \dot{N}^R$ as
\begin{align}
{\dot{E}^R\over \dot{N}^R} 
&= \frac{j^R_E(x+\Delta x)-j^R_E(x)}{j^R(x+\Delta x)-j^R(x)} \nonumber \\ 
&= \mu + {\pi^2 T\over 3} 
\frac{\delta\tau(x+\Delta x)-\delta\tau(x)}{\delta\mu^R(x+\Delta x) - \delta\mu^R(x)} ,
\label{appernr}
\end{align}  
where we used the distribution function (\ref{dfus}) to calculate the current differences in terms of $\delta\tau$ and $\delta\mu^{R}$. 
The first contribution $\mu$ is the energy carried by 
the electron making its transition from the subsystem of right-
to that of left-movers.
The second contribution represents the energy 
of  excitations created at the right Fermi point
during the sequence of three-particle scattering processes that
ultimately results in the backscattering of a right-mover.  
This contribution can also be viewed as the heat transferred from the right-moving subsystem for each backscattering process, in which case, the prefactor $\pi^2 T/3$ is readily obtained from the thermal conductance (\ref{eq:thermal_conductance_noninteracting})  and the relation (\ref{jR_muR}) between right-movers current and chemical potential.

Substituting the ratio of changes in temperature and chemical potential as given by Eq.~(\ref{full_tau}) into Eq.~(\ref{appernr}), one has
\begin{align}
\label{ernrdot}
\frac{\dot{E}^R}{\dot{N}^R}
= - \mu \left[ 1 + {\cal O}\left(\frac{T}{\mu}\right)^2\right],
\end{align} 
where the term in ${\cal O}\left(T/\mu\right)^2$ is discarded in the text, as it only leads to subleading corrections to the conductance. We also briefly mention 
that Eq.~(\ref{ernrdot}) was derived for a 
quadratic dispersion, but can be generalized to the case $\epsilon_p\propto |p|^s$, yielding 
\begin{align}
\frac{\dot{E}^R}{\dot{N}^R} = (1-s) \mu ,  
\end{align} 
again, up to subleading corrections in $T/\mu$.

\end{appendix}


\begin{thebibliography}{99}
\bibitem{footnote} This statement is valid only in the linear response
 limit, when $j,j_Q\to0$, and therefore $\mu$ takes its unperturbed
 position-independent value.
\end{thebibliography}
\end{document}